\documentclass[fleqn,10pt]{wlscirep}
\usepackage[utf8]{inputenc}
\usepackage[T1]{fontenc}
\usepackage{comment}

\title{Filling survey gaps in food security monitoring with spatio-temporal additive Gaussian process models}

\author[1+]{Emma Kopp}
\author[2+]{Sahoko Ishida}
\author[3]{Rebecca Leygonie}
\author[4,5]{Francesca Panero}
\affil[1]{Paris-Dauphine University, CEREMADE, Paris, France}
\affil[2]{University of Oxford, Department of Computer Science, Oxford, United Kingdom}
\affil[3]{ETRO, Vrije Universiteit Brussel, Brussels, Belgium}
\affil[4*]{Sapienza University, Department of Methods and Models for Economics, Territory and Finance, Rome, Italy}
\affil[5]{London School of Economics and Political Science, Department of Statistics, London, United Kingdom}

\affil[*]{francesca.panero@uniroma1.it}

\affil[+]{these authors contributed equally to this work}


\begin{abstract}
Ensuring food security across all regions of a country requires continuous monitoring, yet household surveys often leave significant spatio-temporal gaps due to resource constraints and operational priorities. In this paper, we propose a spatio-temporal additive Gaussian process model to estimate sub-national food security time series by regions. To address the computational cost of Gaussian process models, we exploit Kronecker structure of the spatio-temporal covariance matrix for scalable inference. We evaluate the proposed approach on food security survey data from Nigeria and Chad comparing it against other statistical and machine learning models and show how our proposal achieves better accuracy while retaining reliable uncertainty, especially when covariates are informative. We further apply the model to generate estimates for Nigerian states not covered by the survey, demonstrating its operational value for filling geographic gaps in food security monitoring.\\


\end{abstract}
\begin{document}

\flushbottom
\maketitle
%
%
\thispagestyle{empty}

\section*{Background}

Ensuring food security—the condition in which all people have consistent physical, social, and economic access to sufficient, safe, and nutritious food—remains one of the most pressing global challenges. According to recent estimates, between 638 and 720 million people worldwide, representing 7.8–8.8\% of the global population, faced hunger in 2024 \cite{unicef2025brief}. The threat to food security is particularly acute in regions where conditions can deteriorate rapidly in response to multiple, often interacting stressors such as conflicts\cite{martin2019food,bruck2019reprint}, adverse climate effects\cite{chavez2015end,hasegawa2021extreme} and economic shocks\cite{drammeh2019determinants}. In such volatile contexts, continuous monitoring systems are essential to provide timely situational awareness, enable early warning, and guide the efficient allocation of humanitarian resources.
Household surveys remain one of the most established approaches to measuring food security and an increasingly important tool for informing the efforts of governments and international organisations to address hunger and malnutrition \cite{barrett2010measuring}. 
Nevertheless, sustaining high-frequency surveys with adequate sample sizes across sub-national units is costly and under constrained resources, operational efforts are often prioritised towards areas assessed to be at higher risk, while regions considered relatively stable may be surveyed less frequently or not at all \cite{WFP2021RTM}. 
Although such prioritisation is operationally justified, it introduces gaps in the coverage of observed food security indicators which become problematic when conditions evolve unexpectedly or when retrospective analysis of trends is required. These challenges are likely to intensify in the context of declining humanitarian funding and increasing demands on monitoring systems \cite{GHA2025}.

Machine learning (ML) methods that integrate food security survey data with auxiliary information have been used to estimate and predict sub-national food security indicators. Within this literature, a few studies based on food security data from the mobile Vulnerability Analysis and Mapping (mVAM) program\cite{WFP2021mVMA,mock_morrow_2015_mvam_review, robinson_obrecht2016_mVAM} have been proposed. Extreme Gradient Boosting (XGBoost) has been used to nowcast current prevalence and forecast future conditions when survey observations are unavailable by combining historical survey data with auxiliary covariates \cite{martini2022machine, foini2023forecastability}. Another work has compared a broader set of statistical, machine learning and deep learning models\cite{herteux2024forecasting} on similar data. While these studies highlight the predictive value of auxiliary data, they predominantly emphasise point prediction, with limited attention to uncertainty quantification, despite the clear importance of providing confidence intervals with suitable coverage when reporting estimates in such delicate circumstances. Similarly, while ML methods focus on the prediction accuracy, they usually do not exploit the power of a classical statistical model, which for example properly integrates spatial and temporal information without treating them as simple covariates.\\

In this paper, we address the problem of estimating sub-national food security trends when survey coverage is spatially and temporally incomplete, with a special focus on uncertainty quantification. We consider the Food Consumption Score (FCS), one of the key indicators used by international organisations such as the World Food Program (WFP)\cite{WFP2024FCS}. Our primary outcome of interest is the prevalence of households with insufficient food consumption at the first administrative level. We focus on settings in which some regions within a country are not covered by the data collection, but estimates of regional prevalence trajectories are nevertheless required for monitoring and planning purposes.

Our approach differs from much of the existing literature through the use of a fully Bayesian framework that jointly models spatial and temporal structures and provides reasonable uncertainty quantification. We adopt Gaussian process (GP) models\cite{williams1996gaussian}, which are well suited to spatio-temporal data \cite{cressie2015statistics}. To capture heterogeneous temporal dynamics across space, we employ an additive GP formulation\cite{duvenaud2011additive, lu2022additive} that decomposes variations into spatial effects, temporal effects, and their interaction \cite{ishida2025hierarchical,gelman1995bayesian}, reflecting the assumption that food security trajectories evolve differently across locations due to heterogeneous exposure to shocks and contextual factors. In addition, we incorporate a potentially large set of auxiliary covariates. 

A practical challenge in applying GP models is their computational cost, which scales cubically with the number of observations. We address this by using a Kronecker-based computational strategy \cite{wilson2014fast,pmlr-v37-flaxman15,saatcci2012scalable} that enables scalable inference for spatio-temporal processes defined over fixed spatial units observed across time. We propose a targeted methodological contribution that allows to extend current methods in a setting with random effects. Related work has applied Bayesian spatio-temporal GP models to food security outcomes \cite{bofa2024bayesian}, but our focus differs in its emphasis on additive space–time structure, scalable computation, and uncertainty-aware estimation for regions without survey coverage. Furthermore, instead of yearly-frequency data, we predict at a weekly scale.

We apply the proposed model to Nigeria and Chad, two countries experiencing severe and persistent food security challenges driven by the interaction of conflicts, climate shocks, population displacement and economic stress \cite{GRF2024}.
Nigeria provides a particularly relevant operational setting, as resource constraints and prioritisation mean that some states are surveyed infrequently or not at all, despite the potential for rapid changes in food security conditions. Chad serves as an additional case study characterised by widespread vulnerability, although less stark than Nigeria's.


\section*{Materials and Methods}
This section begins with a description of the datasets used in the analysis. The data are presented by category, with a sector-by-sector overview of the different groups of variables. We then introduce our Gaussian process model, which remains relatively underutilised in this context and constitutes the core of our approach. More technical details, in particular for the computational approach which constitutes a methodological novelty of this paper, are postponed to the Supplementary Materials. Finally, we provide a brief description of the alternative models implemented for comparative evaluation.

\subsection*{Data}

The unit of analysis is region–week at the first administrative level, corresponding to states in Nigeria and provinces in Chad. Therefore,
our target variable is the regional prevalence of households with insufficient food consumption, defined as the proportion of surveyed households whose Food Consumption Score\cite{wfp2024_food_consumption_score} (FCS) falls below a specified threshold.
FCS is a composite indicator reflecting household dietary diversity, frequency of food consumption and the relative nutritional importance of different food groups consumed over the previous seven days. It is important to note that household responses are aggregated and released over a predefined analysis window and reported at a specified administrative level. In Nigeria and Chad, they are aggregated over a rolling 90-day window at the first administrative level. Note that, although data are released with daily frequency, due to the high level of noise in the raw daily estimates, we conduct the analysis at a weekly frequency. For more details on the construction of FCS, we refer to the Supplementary Materials Section 1.

Notation-wise, we will identify $p_{ij} \in [0,1]$ as the prevalence of insufficient food consumption in region $i = 1, \dots, S$ at week $j = 1, \dots, T$. We focus on observations between October 2022 and December 2023, yielding 1575 region–week observations for Nigeria and 1386 for Chad.\\

Covariates are grouped into spatio-temporal, climate and environmental, conflict, economic and demographic variables. A complete list is provided in Table \ref{tab:covariates}. Given the large number of candidate predictors, including all variables was computationally impractical and risked overfitting. We therefore selected a subset of covariates prior to model fitting  for models involving auxiliary variables, with choices driven by evidence from the literature, domain expert knowledge from WFP and model accuracy. 
 
\begin{table}[t!]
\centering
\small
\renewcommand{\arraystretch}{1.1}
\caption{List of inputs variables used in the analysis. Frequencies can be static, monthly (m), dekadal (10d), weekly (w) and daily (d).}
{\begin{tabular}{lccc}
\toprule
\textbf{Variable} & \textbf{Nigeria} & \textbf{Chad} & \textbf{Frequency} \\
\midrule
\multicolumn{4}{l}{\textit{Spatio-temporal data}} \\
\quad Time index & \checkmark &  & w \\
\quad Easting & \checkmark & \checkmark & Static \\
\quad Northing & \checkmark & \checkmark & Static \\ 
\midrule
\multicolumn{4}{l}{\textit{Climate and environmental data}} \\
\quad Rainfall 3-month anomaly & \checkmark &  & 10d \\
\quad NDVI value & \checkmark & \checkmark & 10d \\
\quad NDVI anomaly & \checkmark &  & 10d \\ 
\midrule
\multicolumn{4}{l}{\textit{Conflict data}} \\
\quad Number of Conflicts (Battles) & \checkmark &  & w \\
\quad Fatalities (Battles)  &  \checkmark & & w \\ 
\quad Fatalities (Explosions/bombings)  &  \checkmark & & w \\ 
\midrule
\multicolumn{4}{l}{\textit{Economic data}} \\
\quad Food inflation & \checkmark & \checkmark & m \\
\quad Headline inflation & \checkmark &  & m \\
\quad Exchange rate & \checkmark & \checkmark & d \\ 
\midrule
\multicolumn{4}{l}{\textit{Demographic data}} \\
\quad Administrative area & \checkmark &  & Static \\
\quad MPI & \checkmark &  & Static \\
\quad MPI intensity &  & \checkmark & Static \\
\quad Waterways &  & \checkmark & Static \\
\bottomrule
\end{tabular}}
\label{tab:covariates}
\end{table}

\noindent\textbf{Spatio-temporal variables.}~We include the week index and the geographic coordinates (centroids) of each first administrative region. In the GP framework, these variables define the spatio-temporal input domain and are modelled through the structured priors. In the other models, they are included directly as input features. 

\noindent\textbf{Climate and environmental variables.}~
Environmental indicators consist of rainfall and the Normalized Difference Vegetation Index (NDVI) \cite{liu2016agricultural}. We include both NDVI levels and percentage anomalies relative to a long-term reference period (3 months, to match the rolling time window used to produce the FCS indicator). For more details on the construction of these variables, see the Supplementary Materials Section 1.

\noindent\textbf{Conflict variables.}~
Conflict data are sourced from the ACLED database \cite{raleigh2010introducing}. For each region and week, we compute rolling 90-day sums of the number of events and associated fatalities for three event types: battles, violence against civilians, and explosions or remote violence. Fatalities were log-transformed (substituting zeros with ones).

\noindent\textbf{Economic variables.}~
Economic indicators include log-transformed headline inflation, food inflation and exchange rates. They are recorded daily by WFP publicly available Economic Explorer Inflation rates and aggregated to weekly frequency. These variables are defined at the national level and therefore take the same value across all regions at a given time point.

\noindent\textbf{Demographic and structural variables.}~
Time-invariant regional characteristics include the logarithm of the area size, Muslim population share, the Waterways Index \cite{foini2023forecastability} and the Multidimensional Poverty Index (MPI) \cite{alkire2025global}.\\

All covariates were aligned to weekly frequency. For variables originally recorded at monthly resolution, we applied cubic spline interpolation to obtain weekly values. All covariates were standardized prior to model estimation. Interaction terms were created between 
food inflation and MPI, and the logarithm of food inflation and MPI, to capture differential effects of price shocks across poverty levels.

\subsection*{Additive Gaussian process model}
Each target observation is associated with covariates $\mathbf{x}_{ij}\in\mathbb{R}^d$, a spatial location $\mathbf{s}_i\in\mathbb{R}^2$ corresponding to the centroid of region $i$, and a temporal coordinate $t_j\in\mathbb{R}$, measured in weeks. We model the logit-transformed prevalence, 
$
y_{ij} = \log\left(p_{ij}/(1-p_{ij})\right),
$
as
\begin{equation*}
y_{ij} \sim \mathcal{N}(\mu_{ij}, \sigma_y^2),
\end{equation*}
where $\sigma_y^2$ denotes the measurement error variance and the latent mean is specified as
\begin{equation*}
\mu_{ij} = \beta_0 + \mathbf{x}_{ij}^\top \boldsymbol{\beta} + f(\mathbf{s}_i, t_j) + u_i,
\end{equation*}
where $\beta_0$ is a global intercept representing the average prevalence (in logit scale) across regions and time, $\boldsymbol{\beta}$ is a vector of regression coefficients associated with the covariates, $f(\mathbf{s}_i, t_j)$ captures the spatio-temporal dependence, and $u_i$ denotes an unstructured region-specific random effect accounting for spatially uncorrelated heterogeneity, which we assume to follow $N(0,\sigma_u^2)$. 

To flexibly model nonlinear spatio-temporal structures, a Gaussian process \cite{rasmussen2006gaussian} is a natural choice. Specifically, we place an additive GP \cite{duvenaud2011additive, lu2022additive, ishida2025hierarchical} prior on the function $f(\mathbf{s}_i, t_j)$ of the form
\begin{equation*}
f(\mathbf{s}_i, t_j) = f_s(\mathbf{s}_i) + f_t(t_j) + f_{st}(\mathbf{s}_i, t_j),
\end{equation*}
where $f_s(\mathbf{s}_i)$ and $f_t(t_j)$ represent the main effects of space and time, respectively, and $f_{st}(\mathbf{s}_i, t_j)$ captures their interaction.
The main effects are modelled as independent zero-mean Gaussian processes,
$
f_s \sim \text{GP}(0, k_s), f_t \sim \text{GP}(0, k_t),
$
where $k_s$ and $k_t$ are covariance kernels defined on $\mathbb{R}^2\times\mathbb{R}^2$ and $\mathbb{R}\times \mathbb{R}$, respectively. The interaction term is also modelled as a zero-mean Gaussian process,
$f_{st} \sim \text{GP}(0, k_{st}),
$
with covariance function given by the product kernel $k_{st} = k_s \otimes k_t$, where $\otimes$ is the tensor product of the covariance functions, a standard choice for interaction terms. 
For the spatial main effect, we use an exponential covariance kernel
\begin{equation*}
k_s(\mathbf{s},\mathbf{s}')=\alpha_s^2\exp\!\left(-\frac{\lVert \mathbf{s}-\mathbf{s}'\rVert}{\rho_s}\right),
\qquad \mathbf{s},\mathbf{s}'\in\mathbb{R}^2,
\end{equation*}
where $\alpha_s^2$ controls the marginal variance and $\rho_s$ the spatial length-scale.
For the temporal main effect, to take into account the rougher trajectories of FCS over time and the non-stationarity of the process, we use a sum of a stationary and a non-stationary component
$k_t(t,t') = k_{t1}(t,t') + k_{t2}(t,t'),\,t,t'\in\mathbb{R}$.
The stationary component is also taken to be exponential,
\begin{equation*}
k_{t1}(t,t')=\alpha_{t1}^2\exp\!\left(-\frac{|t-t'|}{\rho_{t1}}\right).
\end{equation*}
For the non-stationary component, we follow \cite{bergsma2020regression} and use a squared Brownian motion kernel. Specifically, we start from the fractional Brownian motion (fBM) kernel and set the Hurst coefficient to $\gamma=0.5$ (corresponding to standard Brownian motion), yielding
\begin{equation*}
k_{\mathrm{BM}}(t,t') = \frac{\alpha_{t2}^2}{2}\Big(|t| + |t'| - |t-t'|\Big).
\end{equation*}
We then construct the squared kernel as $k_{t2}(t,t')=\sum_{\ell=1}^{T} k_{\mathrm{BM}}(t,t_\ell)\,k_{\mathrm{BM}}(t',t_\ell),$
This squared construction yields smoother sample paths than the standard Brownian motion kernel\cite{bergsma2020regression}.

Posterior inference is performed using Markov chain Monte Carlo, with weakly informative hyperpriors placed on covariance parameters. Given the GP prior on the latent function, both the posterior and the posterior predictive distribution remain GPs. To handle the computational burden inherent to large spatio-temporal datasets, we exploit the separable structure of the covariance function: Kronecker-based methods \cite{wilson2014fast, saatcci2012scalable} reduce the cost of matrix operations substantially, and have been extended to additive GPs with interaction terms\cite{ishida2025hierarchical}. 
In our setting, we need to extend this method to include the unstructured spatial random effect. Full derivations and implementation details are provided in the Supplementary Material Section 3.\\

To evaluate the performance of the proposed spatio-temporal GP model, we implemented four alternative approaches for comparison: a GP model without covariates (i.e. a regression with mean $\tilde{\mu}_{ij}=\beta_0 + f(\mathbf{s}_i, t_j) + u_i$), a Bayesian Ridge regression, 
a multilayer perceptron (MLP), and XGBoost. The models are implemented on the same set of covariates but, differently from the GP where space and time are structurally included in the model, in these comparative models space and time are both used as covariates since data are treated as independent observations rather than time series. For details on the individual implementation, we refer to the Supplementary Materials Section 4.

It is noteworthy that XGBoost is currently used operationally within WFP for nowcasting and short-term forecasting \cite{martini2022machine, foini2023forecastability}, and predictions are displayed on the WFP platform HungerMap\cite{WFPHungerMapLive}. This comparison provides a benchmark against a widely used operational model. 

\section*{Results}

Model performance in predictive accuracy was evaluated using Mean Absolute Error (MAE) and root Mean Squared Error (RMSE). To evaluate uncertainty quantification, we compute the mean interval length (MIL) of 95\% predictive intervals and their empirical coverage. Since MLP and XGBoost do not natively provide principled uncertainty quantification, prediction intervals are constructed using Monte Carlo dropout and bootstrap aggregation, respectively. In contrast, the Bayesian GP framework naturally propagates parameter and latent process uncertainty throughout the model, resulting in more coherent and better calibrated predictive intervals.

Since a classic time series cross validation scheme would not be apt for our goal, i.e. to fill survey gaps by predicting FCS on unseen regions, all metrics were averaged over a spatially balanced cross-validation scheme: we created folds that are balanced across different geographical regions, with five folds for Nigeria and six folds for Chad. Hyperparameters of the various models were tuned using validation sets inside the same scheme. A complete account of the cross validation and training details can be found in the Supplementary Materials Section 2. 

\begin{table}[t!]
    \centering
    \caption{Metrics results for Nigeria and Chad across methods}
    \label{tab:prediction_metrics}
    \begin{tabular}{llcccc}
    \toprule
    & & & & \multicolumn{2}{c}{95\% CI} \\
    \cmidrule(lr){5-6}
    Country & Model & RMSE & MAE & Coverage & MIL \\
    \midrule
    Nigeria & Bayesian Ridge & 0.052 & 0.045 & 70.3\% & 0.112  \\
            & GP without covariates & 0.074 & 0.069 & 100 \% & 0.476 \\
            & GP with covariates & 0.047 & 0.042 & 99.4\% & 0.339 \\
            & MLP & 0.065 & 0.056 & 74.5\% & 0.179 \\
            & XG-Boost &  0.057 & 0.050 & 25.2\% & 0.037 \\ 
    \midrule
    Chad & Bayesian Ridge & 0.065 & 0.055 & 60.1\% & 0.116 \\
         & GP without covariates & 0.053 & 0.046 & 96.6\% & 0.287 \\
         & GP with covariates & 0.052 & 0.047 & 96.8\% & 0.294 \\
         & MLP & 0.053 & 0.045 & 73.1\% & 0.133 \\
         & XG-Boost & 0.059 & 0.053 & 19.0 \% & 0.030 \\
    \bottomrule
    \end{tabular}
\end{table} 

\subsection*{Validation in Nigeria} 
For Nigeria, the GP model with covariates and the Bayesian Ridge regression model were fitted using the subset of auxiliary predictors described in table \ref{tab:covariates}.
A linear time index (day number) was also incorporated. Although nonlinear temporal dynamics are captured by the temporal Gaussian process component, the linear term allows the model to account for a systematic monotonic trend over the study period (which, especially for Nigeria, is seen in Figure 5 in the Supplementary Materials).


In terms of predictive accuracy, the GP model with covariates achieves the lowest error among all competing approaches (Table~\ref{tab:prediction_metrics}), improving upon all other methods. Regarding uncertainty quantification, the GP credible intervals achieve an empirical coverage exceeding the nominal level. Although conservative, this coverage is substantially more reliable than that of MLP, Bayesian Ridge and XGBoost, whose intervals are markedly shorter and under-calibrated. 

\begin{figure}[t!]
    \centering
    \includegraphics[width=0.99\linewidth]{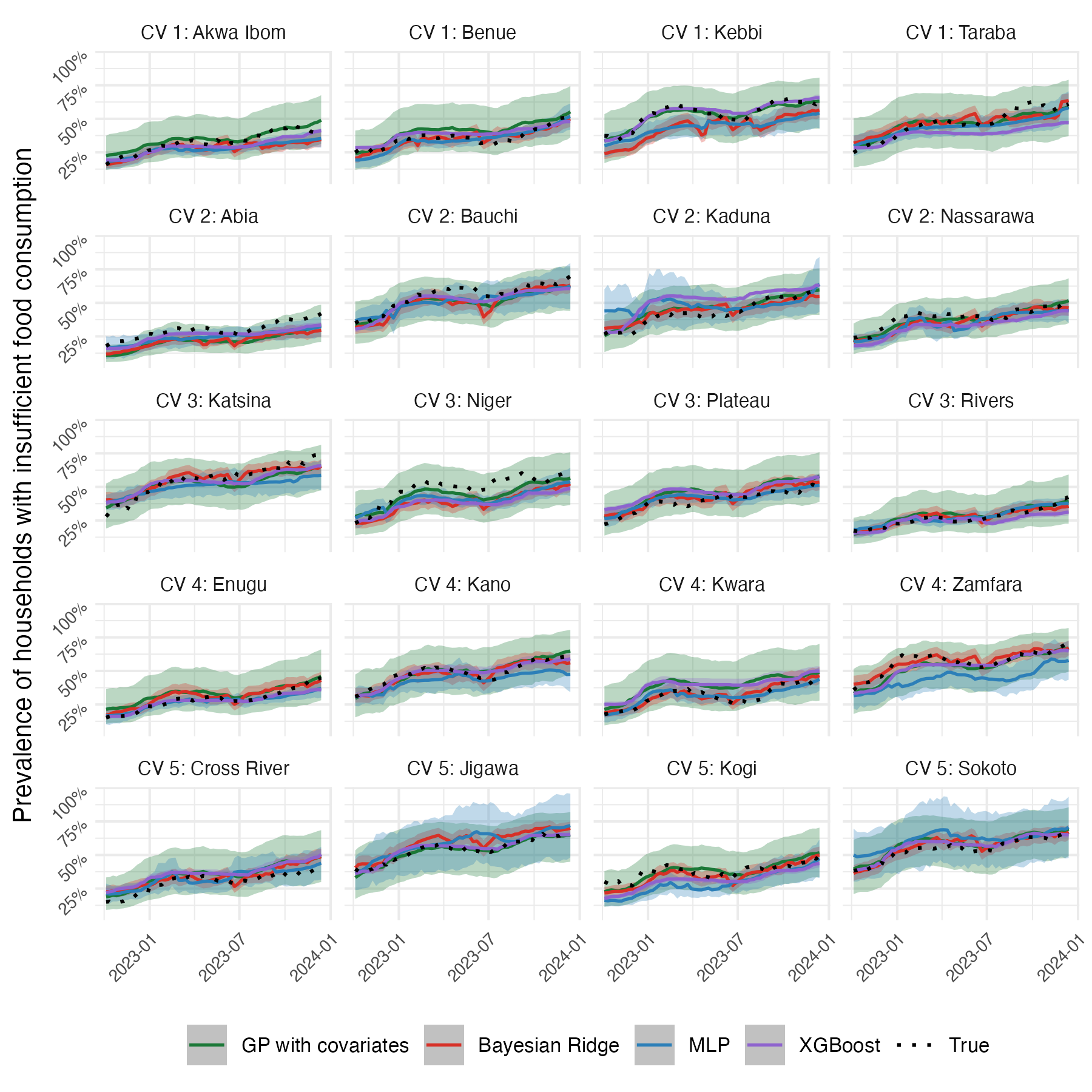}
    \caption{Prevalence of households with insufficient food consumption across states in Nigeria. The black dotted line denotes the observed survey-based prevalence. Coloured solid lines correspond to predictions from Gaussian Process (GP) with covariates, Bayesian Ridge, MLP, and XGBoost models. Shaded regions represent 95\% predictive intervals.}
    \label{fig: NGA prediction all regions}
\end{figure}

\begin{figure}[t!]
    \centering
    \includegraphics[width=0.99\linewidth]{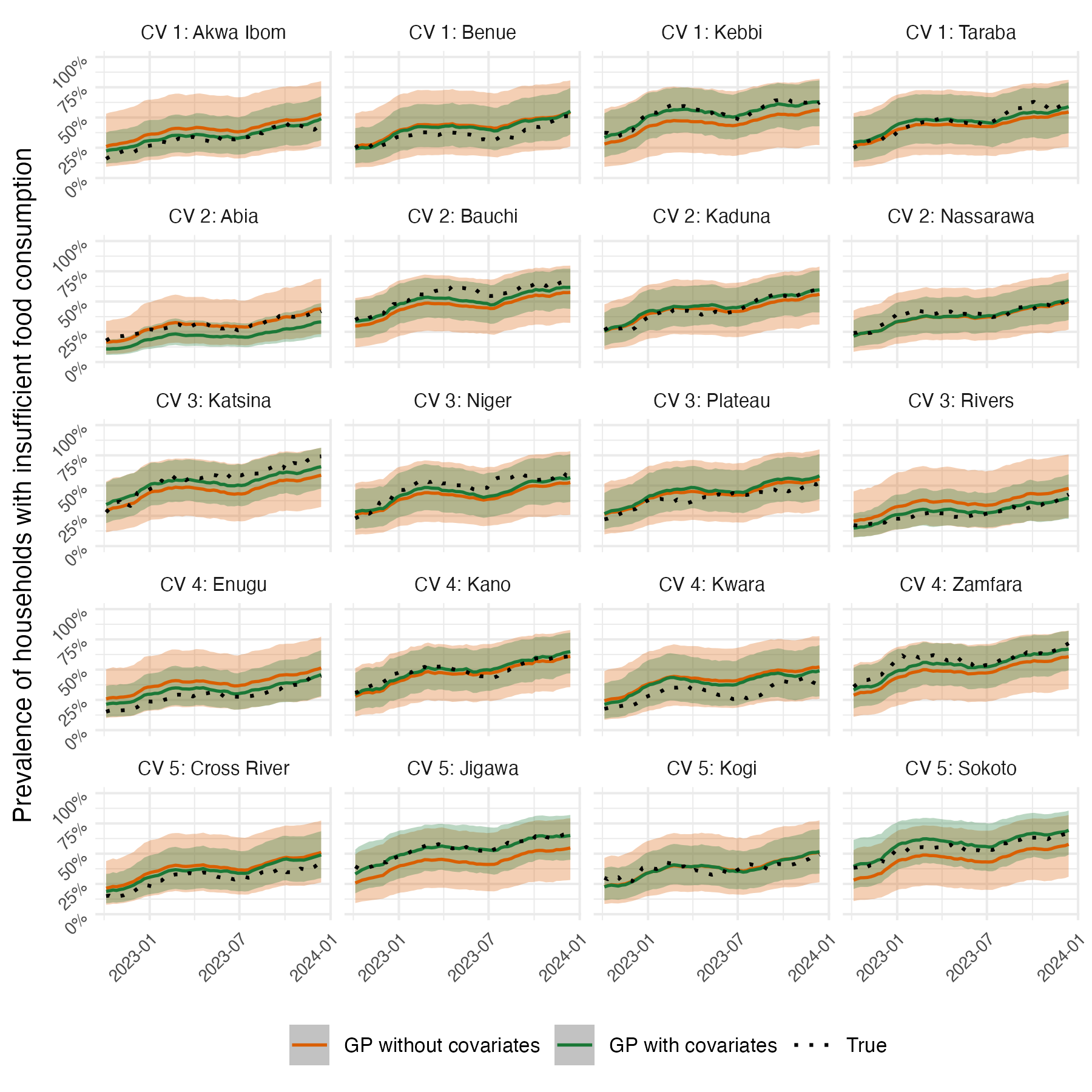}
    \caption{Prevalence of households with insufficient food consumption across states in Nigeria. The black dotted line denotes the observed survey-based prevalence. Coloured solid lines correspond to predictions from Gaussian Process (GP) with covariates, and without covariates. Shaded regions represent 95\% predictive intervals.}
    \label{fig: NGA GP comparison}
\end{figure}

Figure~\ref{fig: NGA prediction all regions} presents the predicted weekly prevalence trajectories of households with insufficient food consumption for each region under the GP with covariates, Bayesian Ridge, MLP and XGBoost models. 
It is interesting to note that the bootstrap method to provide uncertainty for XGBoost results in extremely narrow confidence intervals, with a extremely low coverage, suggesting that other methods should be evaluated to provide a more trustworthy operational setting.
Figure~\ref{fig: NGA GP comparison} compares GP predictions with and without auxiliary covariates for the test states. The inclusion of covariates visibly reduces the width of the predictive intervals ($29\%$ reduction on average) while maintaining coverage, demonstrating the value of auxiliary information in constraining posterior uncertainty. 

The primary focus in this section is predictive performance and uncertainty quantification rather than detailed interpretation of individual regression coefficients; posterior means of the regression parameters are therefore reported in the Supplementary Material Figure 8.

\subsection*{Validation in Chad}
For Chad, the GP model with covariates and the Bayesian Ridge regression model were fitted using a more parsimonious set of auxiliary predictors, which were selected from the same set of Nigeria by retaining the biggest set which allowed sound convergence of the algorithms (Table \ref{tab:covariates}). 
As in the Nigerian specification, a linear time index (day number) was incorporated. 

In terms of predictive accuracy (Table~\ref{tab:prediction_metrics}), the GP model with covariates achieves an RMSE and MAE closely matching the MLP and outperforming Bayesian Ridge and XGBoost. While predictive accuracy is comparable between GP and MLP, the difference becomes clear when evaluating uncertainty quantification. The GP credible intervals achieve empirical coverage of $96\%$, very close to the nominal level. In contrast, Bayesian Ridge, MLP and XGBoost under-calibrate with substantially shorter intervals. Incorporation of the auxiliary information in the GP did not lead to clear improvement in predictive accuracy but provided reduction in prediction interval width of $14\%$. The visual performance of the various methods is illustrated in Figure 9 of the Supplementary Materials. Posterior means of the regression coefficients are reported in the Supplementary Material Figure 8.

\subsection*{Real-world application to regions not covered by the survey in Nigeria}

A key distinction of this study relative to existing work employing XGBoost\cite{martini2022machine, foini2023forecastability} and Reservoir Computing\cite{herteux2024forecasting} lies in both the objective and the modelling approach. The aim here is to estimate the trajectory of food insecurity indicators in regions where no survey data are available over the entire study period, rather than to nowcast or forecast values in regions with observed data and extend predictions beyond the observed time horizon. Our setting requires prediction at unobserved spatial locations by borrowing information from surveyed regions, rather than interpolation within observed regions or short-term temporal forecasting, and relies on known, observed levels of covariates rather than values forecasted into the future. From a methodological perspective, the spatio-temporal GP model is well suited to this objective, as it explicitly models spatial and temporal dependence and enables information to be shared across regions and time. With such an approach, credible intervals are provided for free for model parameters, which is not possible for the competing frequentist models.

We now estimate the GP model with covariates, previously selected as the best performing model in our analysis for Nigeria, on the 25 states with available survey data, and subsequently produce predictions for the states not covered by the mVAM survey. These regions are generally considered to be in relatively stable conditions and are therefore not routinely prioritized for data collection. This setting reflects a realistic operational scenario in which model-based estimates are required to fill geographic gaps in survey coverage.

\begin{figure}
    \centering
    \includegraphics[width=0.99\linewidth]{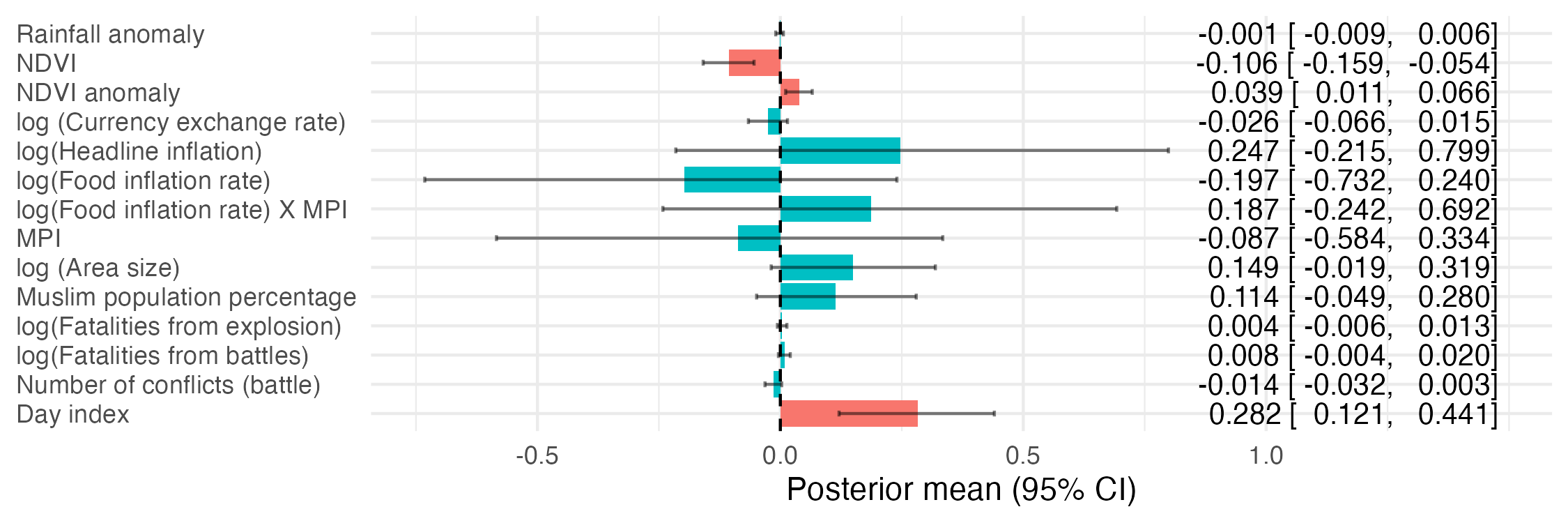}
    \caption{Posterior means and 95\% credible intervals from the GP model with covariates. Bars represent posterior means and horizontal lines denote 95\% credible intervals. Variables shown in red have credible intervals that do not include zero.}
    \label{fig: nga reg coef summary}
\end{figure}

Figure~\ref{fig: nga reg coef summary} presents posterior means and 95\% credible intervals for the regression coefficients, which indicates negative association for NDVI and a positive association for NDVI anomaly, as their credible intervals exclude zero. Substantively, this suggests that while higher baseline vegetation levels are associated with lower values of the outcome, deviations from typical vegetation conditions are positively related to it. In addition, the day index shows a statistically significant positive effect, capturing a linear (global) time trend in the outcome over the study period.

\begin{figure}[t!]
  \centering
  \begin{subfigure}[b]{0.99\linewidth}
    \centering
    \includegraphics[width=\linewidth]{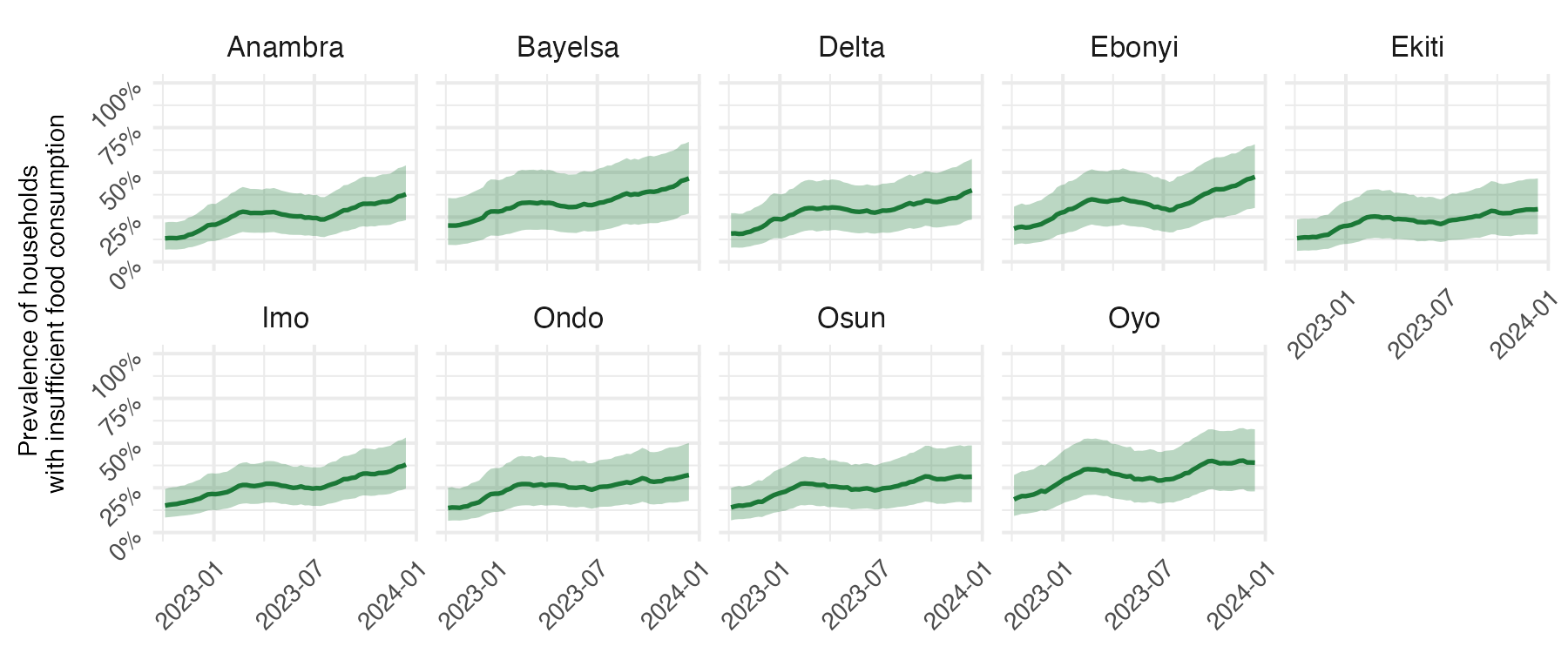}
    \caption{Prediction for unsurveyed regions}
    \label{fig:nga_prediction_unsurveyed}
  \end{subfigure}
  \begin{subfigure}[b]{0.95\linewidth}
    \centering
    \includegraphics[width=\linewidth]{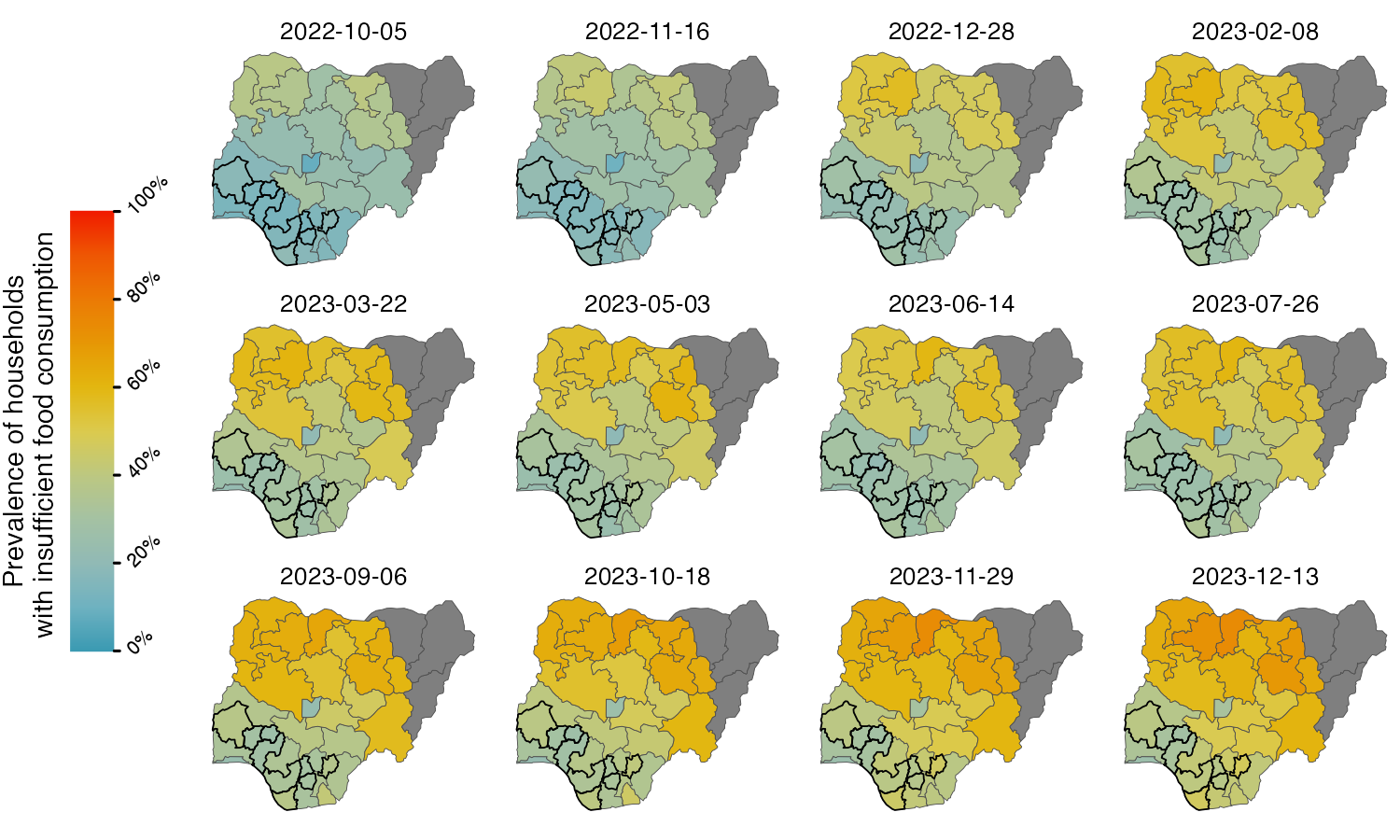}
    \caption{Spatial distribution }
    \label{fig:nga_spatial_distribution}
  \end{subfigure}
  \caption{(Top) Estimated prevalence of households with insufficient food consumption over the study period for states not covered by the mVAM survey. Solid green lines represent predictions from a Gaussian Process (GP) model with covariates, and shaded areas indicate 95\% confidence intervals. (Bottom) Spatial distribution of the prevalence of households with insufficient food consumption across Nigeria over time. States outlined with a solid black border denote regions not covered by the mVAM survey, where values are GP-based estimates. In all other states, values correspond to observed mVAM survey data.}
 \label{fig:NGA unsurveyed regions}
\end{figure}

Figure~\ref{fig:nga_prediction_unsurveyed} top panel presents the estimated prevalence of households with insufficient food consumption over the study period for the unsurveyed states. Prevalence levels are relatively moderate at the beginning of the study period, around 20\%, followed by a noticeable increase in early to mid-2023 and a continued upward trend toward the end of 2023. This overall trajectory mirrors the national pattern documented in surveyed states.
While most unsurveyed regions follow this general upward trend, there is meaningful heterogeneity in their evolution. Some states exhibit relatively flat trajectories, such as Ekiti, while other states show more pronounced deterioration. Several states, including Ebonyi and Bayelsa, are estimated to exceed the 40\% threshold on average at certain points, indicating a shift toward very high levels of insufficient food consumption\cite{WFPHungerMapLive}. Regarding uncertainty, we observe a similar heterogeneity across regions. For most unsurveyed states, the predictive intervals are relatively narrow, with upper bounds of the 95\% credible intervals remaining below 50\%. However, several states such as Bayelsa, Ebonyi and Oyo, exhibit wider predictive intervals and upper bounds reaching approximately 60\%. 

Figure~\ref{fig:nga_spatial_distribution} bottom panel shows the corresponding spatial distribution over time. States outlined with a solid black border denote those not covered by the mVAM survey. For these regions, values are GP-based model estimates, while in all other states, values correspond to observed mVAM survey data. Spatially, the predicted values in unsurveyed regions are broadly consistent with conditions in neighbouring states. In general, estimated prevalence levels align with surrounding regions and evolve in parallel with broader national dynamics, with higher prevalence observed across much of the country during 2023.

It is important to emphasize that estimates for unsurveyed states are entirely model-based and cannot be directly validated against mVAM survey data. As such, these results should be interpreted cautiously. The strengths and limitations of relying on model-based predictions in the absence of ground-truth observations are discussed further in the Discussion section.

\section*{Discussion}

In this study, we demonstrate the use of spatio-temporal GP regression to estimate the prevalence of households with insufficient food consumption, a key food security indicator used operationally by WFP, in regions where surveys are not conducted. The model is implemented in a fully Bayesian framework, allowing uncertainty to propagate through all stages of model estimation and prediction. Uncertainty, in fact, plays a particularly important role in such a delicate application and is critical in data-sparse settings. By modelling spatial and temporal dependence, the approach enables the estimation for unsurveyed regions to be informed by patterns observed in neighbouring regions and by the overall temporal dynamics in the data. We also evaluate whether incorporating auxiliary variables describing environmental, economic, demographic, and conflict-related conditions can further improve predictive accuracy and the calibration of prediction intervals. Model performance is assessed through cross-validation experiments in two countries facing recurrent food security challenges, Nigeria and Chad, followed by a real-world application in which the model is used to generate estimates for Nigerian regions where no survey observations are available. Other than the applied results, we propose a methodological contribution in the form of an extension of present Kronecker decompositions for the Gram matrix of a GP regression to allow for country specific random effects. 

The spatio-temporal GP model with covariates achieves the best overall performance among the models considered. In Nigeria, it improves predictive accuracy relative to competing approaches, with RMSE reductions of at least 10\% compared to the alternative methods. In terms of uncertainty quantification, it is the only approach that achieves coverage above the nominal 95\% level (99.4\%), whereas alternative models exhibit substantial under-coverage. Although the resulting credible intervals are moderately conservative, incorporating covariates reduces by $14\%$ their length relative to the GP model without covariates, indicating improved precision without compromising reliability. In Chad, the GP model with covariates also ranks among the best-performing models in terms of predictive accuracy, achieving MAE and RMSE values comparable to or lower than competing approaches. It provides well-calibrated uncertainty estimates, with coverage close to the nominal level (96.8\%), while alternative models again tend to under-cover.

Our results demonstrate that combining spatio-temporal structure with auxiliary variables enables reliable estimation in regions without survey coverage. The contribution is not to replace survey-based assessment, but to provide statistically grounded estimates, together with associated uncertainty, in situations where some regions are not covered due to resource constraints or operational disruptions. Such estimates can support monitoring by identifying regions where predicted conditions suggest potential deterioration, thereby informing whether survey efforts should be prioritised or resumed. In Nigeria, for example, an alert mechanism based on the model output could suggest when to resume data collection in currently unsurveyed states. Considering the classification thresholds for FCS, one operational rule could be to trigger data collection when the upper bound of the 95\% credible interval exceeds the high food insecurity threshold for a predetermined consecutive period.\\

Although the results are encouraging, some caution in interpretation is needed, and various model improvements can be proposed. Many auxiliary variables exhibit spatial and temporal structure similar to the outcome variable, making it difficult to disentangle their effects from the spatio-temporal component of the model. Estimated regression coefficients may therefore reflect shared structure rather than purely marginal effects. Furthermore, our model assumes linear relationships between covariates and the outcome, with only a limited number of interaction terms included. More flexible approaches could be considered, for example combining GP components with boosting-based modelling of covariate effects \cite{sigrist2022gpboost}, or placing a flexible GP prior on the covariate-response function. While such approaches would allow for richer representations, they would also introduce additional computational complexity in the spatio-temporal setting. On the target variable side, our analysis focuses on a single food security indicator. As food security is inherently multidimensional, relying on a single indicator provides only a partial view of underlying conditions. Extending the framework to additional indicators, such as reduced coping strategies or other complementary measures, would be necessary to support a more comprehensive assessment.

The contribution of auxiliary covariates differed markedly between the two countries, and we believe this reflects genuine differences in the nature of food insecurity in each setting. In Nigeria, the food security situation is not only severe but also highly volatile: Nigeria was the second most affected country by food insecurity globally in 2023\cite{GRF2024}, and conditions vary substantially across regions and over time in response to conflict, economic shocks, and climate variability. This heterogeneity creates residual variation that covariates such as NDVI, its anomaly, and the temporal trend index can meaningfully capture, explaining their strong predictive contribution. This is also highlighted by the poor performance of the purely spatio-temporal GP, which performs worst in terms of point metrics and has an unreasonably wide coverage. In Chad, by contrast, food insecurity is more spatially uniform and persistently high across the study period, with the spatio-temporal GP component alone able to capture most of the outcome variation. This leaves little residual signal for auxiliary variables to explain, which accounts for the more limited improvement from covariate inclusion in that setting. These findings highlight that the value of auxiliary information is context-dependent and should be evaluated empirically for each application rather than assumed.

Another reason to opt for country-specific models is to maximise both predictive performance and convergence requirements. In Chad, we included the maximum number of covariates for which satisfactory MCMC convergence diagnostics were achieved, which was probably more difficult due to the lower number of data points compared to Nigeria. Although this requires more operational effort than a unified model, it minimises prediction errors for each country and acknowledges that different countries may have different drivers of food insecurity\cite{FAO2026}. A different modelling direction could be to design a spatial hierarchical model, where covariate effects are specified hierarchically across countries and geographical areas such as West Africa. Though, due to the more complicated structure of such models, we expect convergence issues might arise.

It is also important to acknowledge a limitation of the real-world application to unsurveyed Nigerian states. The absence of survey data in these regions is not random: states were deemed lower priority precisely because conditions were perceived as relatively stable, introducing a form of informative missingness. This means that the cross-validation results, obtained on regions that were surveyed and are therefore more representative of the training data, may overstate predictive performance for the truly unsurveyed states. In practice, the direction of this bias is operationally conservative: if conditions in unsurveyed regions are indeed more stable, the model trained on more volatile surveyed regions will over-estimate uncertainty, which is preferable to under-estimating it in a humanitarian monitoring context. Addressing informative missingness more formally, for example by incorporating data on the survey allocation process itself\cite{diggle2010geostatistical}, represents a worthwhile direction for future work.

Finally, a word on the uncertainty intervals for the machine learning baselines. Those have been constructed using Monte Carlo dropout for the MLP and bootstrap aggregation for XGBoost, the latter being the method currently used in WFP's operational implementation \cite{martini2022machine, foini2023forecastability}. While retaining these methods ensures a practically relevant comparison, both are known to produce poorly calibrated intervals, as confirmed by the substantial under-coverage observed across both countries. More refined approaches to uncertainty quantification, such as conformal prediction or quantile regression, could be explored in future work to provide better calibration within these frameworks\cite{vovk2005algorithmic}.

\bibliography{sample}

\section*{Acknowledgements}
We thank the Early Warning and Forecasting Unit led by Duccio Piovani at World Food Program for the support on the initial idea and the help with data processing.
FP and SI thank the London School of Economics and Political Science for the support received under the Research and Impact Support Fund 2023/2024.

\section*{Author contributions statement}
SI and FP were responsible for conception of the study. FP provided overall supervision for the study, ran XGBoost and Bayesian ridge models. SI was responsible for data pre-processing of Nigeria database. EK was responsible for data pre-processing of Chad database. RL was responsible for running MLP models. SI and EK were responsible for data analysis, the study design and interpretation of the results for Nigeria and Chad and manuscript writing. SI, FP and EK jointly wrote the paper. All authors read and approved the final manuscript.

\section*{Data and code availability} 
Both target and covariate datasets were accessed through WFP’s internal APIs and subsequently processed for analysis. Due to data sharing restrictions, the processed datasets are not publicly available but may be obtained from the authors upon reasonable request and subject to permission from WFP. Part of the target data used in this study are derived from the World Food Programme (WFP) HungerMap platform and can be accessed via the WFP API (see available documentation at \url{https://docs-wfp-hungermap.netlify.app/docs/chatbot/data_retrievals/}). Auxiliary input data were obtained from publicly available sources, including the WFP VAM DataViz portal (\url{https://dataviz.vam.wfp.org/}) and the Humanitarian Data Exchange (\url{https://data.humdata.org/}). Additional demographic data on Muslim population percentages in Nigeria were obtained from published estimates.\cite{nwankwo2019religion}. 
The code used in this study is available in public repository (\url{https://github.com/emmakopp1/efficientGP.git}).

\section*{Competing interests}
The authors declare no competing interests.

\newpage

\section*{SUPPLEMENTARY MATERIALS}

\section{Data}

\subsection*{Target variable}

Within the WFP mVAM programme, a Real-Time Monitoring (RTM) system \cite{WFP2021RTM} is designed for continuous monitoring purposes. enabling the regular production of updated indicators based on rolling survey data. 
Aggregated estimates are computed over a predefined analysis window and reported at a specified administrative level. The choice of administrative unit and time window reflects operational objectives and resource constraints, and determines the required survey intensity.
In Nigeria and Chad, FCS data are aggregated over a rolling 90-day window at the first administrative level. For each region and reporting date, WFP estimates 
the prevalence of insufficient food consumption as the proportion of surveyed households within the corresponding 90-day window classified below the operational threshold. WFP operationally targets at least 150 completed household interviews per region per analysis window, regardless of the window length. For Chad, the average number of households per region–window is 264.2, ranging from 217 to 285. In Nigeria, regions excluding Lagos, Kano, and Abuja have an average of 446.7 households, with values ranging from 421 to 462. In contrast, Lagos, Kano, and Abuja have substantially higher counts, with an average of 896.5 households and a range of 874 to 915. 

\begin{figure}[b!]
\centering
\begin{subfigure}{0.495\textwidth}
    \centering
    \includegraphics[width=\textwidth]{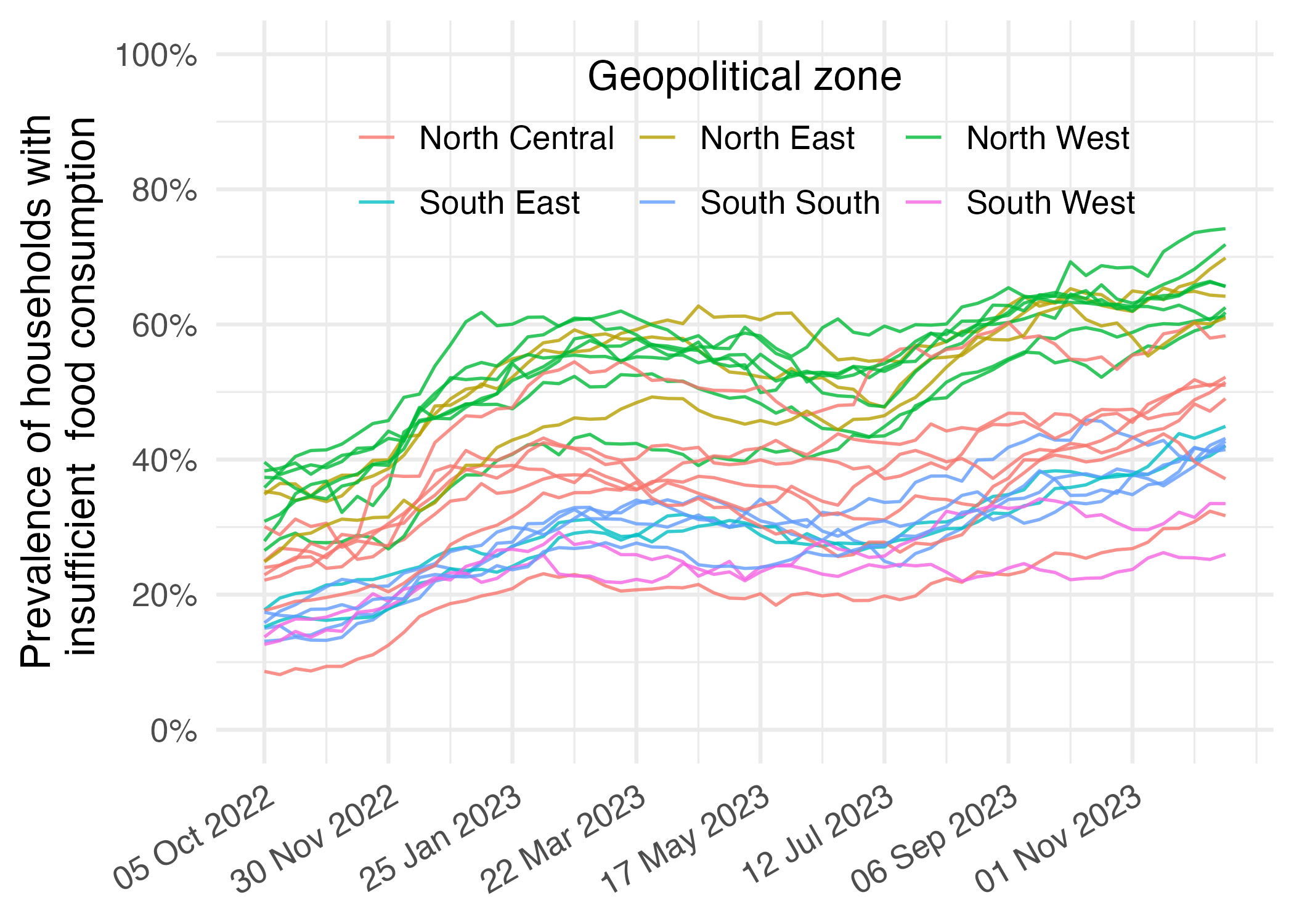}
    \caption{Nigeria}
    \label{fig:nga_fcs_by_region}
\end{subfigure}
\hfill
\begin{subfigure}{0.495\textwidth}
    \centering
    \includegraphics[width=\textwidth]{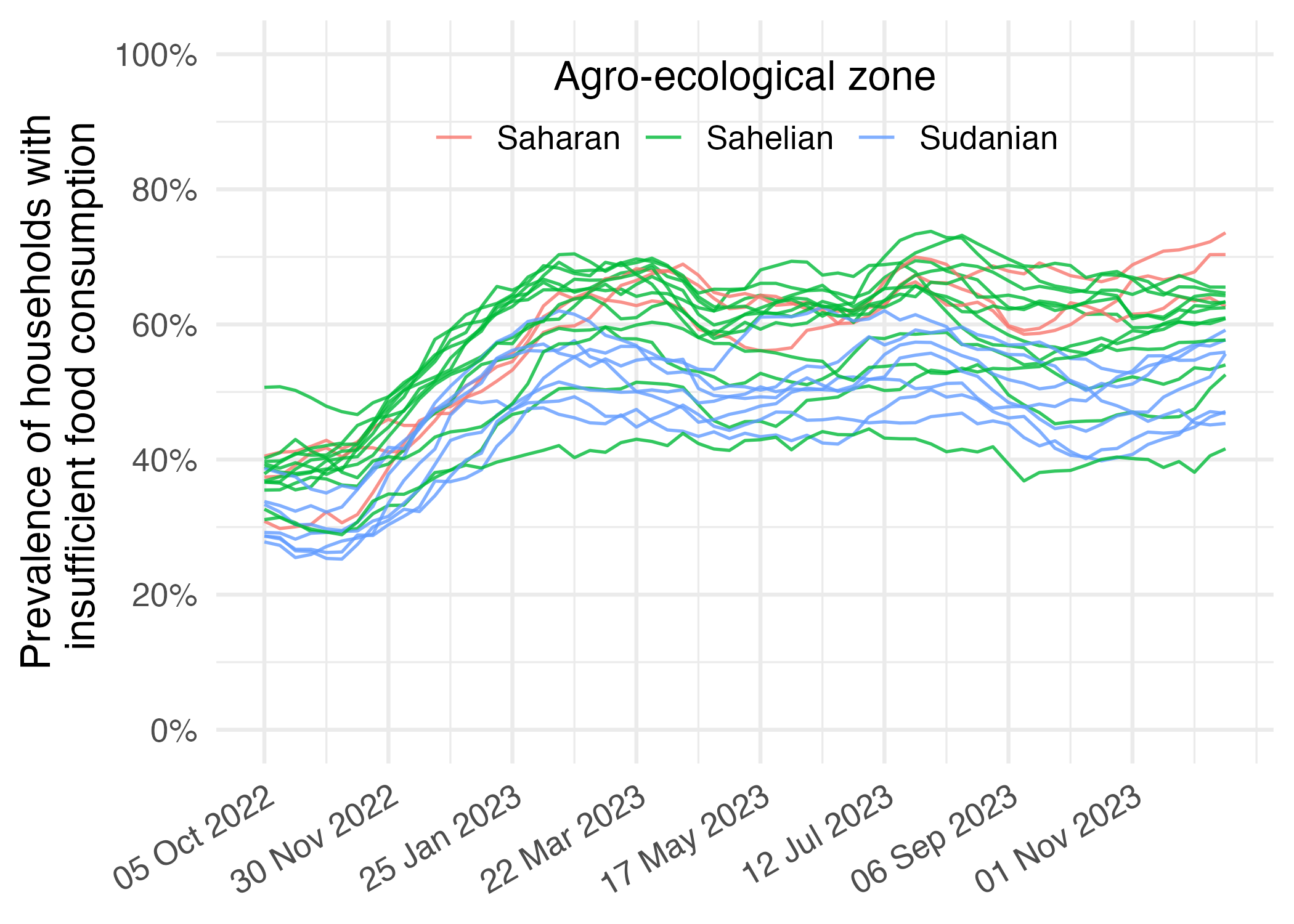}
    \caption{Chad}
    \label{fig:tcd_fcs_by_region}
\end{subfigure}
\caption{Evolution of the prevalence of households exhibiting insufficient food consumption patterns in Nigeria and Chad, shown at the first administrative level (states for Nigeria and provinces for Chad). Lines represent individual administrative regions over time. Colours indicate regional groupings used for visual guidance: geopolitical zones for Nigeria and agro-ecological zones for Chad. All first-level administrative regions are included for Chad. For Nigeria, the analysis includes 25 states, covering all six geopolitical zones: North West (Jigawa, Kaduna, Kano, Katsina, Kebbi, Sokoto, Zamfara), North East (Bauchi, Gombe, Taraba), North Central (Benue, Kogi, Kwara, Nassarawa, Niger, Plateau, Abuja), South West (Lagos, Ogun), South East (Abia, Enugu), and South South (Akwa Ibom, Cross River, Edo, Rivers).
}
\label{fig:fcs_by_region}
\end{figure}

The Food Consumption Score is constructed from survey responses on the consumption frequency of eight standard food groups and ranges from 0 to 112. According to WFP thresholds, households are classified as \textit{poor} if FCS $< 21$, \textit{borderline} if $21 \leq$ FCS $< 35$, and \textit{acceptable} if FCS $\geq 35$. Insufficient food consumption corresponds to FCS lower than 35.
Figures \ref{fig:nga_fcs_by_region} and \ref{fig:tcd_fcs_by_region} show the ground truth evolution of the prevalence of households with insufficient food consumption by region in both countries. For both countries, the prevalence shows an upward trend, reflecting a deterioration in food security throughout the study period.

\subsection*{Climate and environmental variables details}

Rainfall estimates \cite{haile2005weather} are derived from satellite observations combined with rain gauge data and are released at a 10-day frequency (dekadal), and values are spatially averaged over administrative regions. 
NDVI is constructed from MODIS satellite products (MOD13C1 and MYD13C1), combined into 8-day composites and filtered using a Whittaker smoothing procedure to reduce atmospheric noise and fill missing values. The filtered data are interpolated to dekadal frequency and averaged over administrative regions. Dekadal observations are mapped to their start date.

\subsection*{Nigeria Survey Coverage}
\begin{figure}[!htbp]
    \centering
    \includegraphics[width=0.7\linewidth]{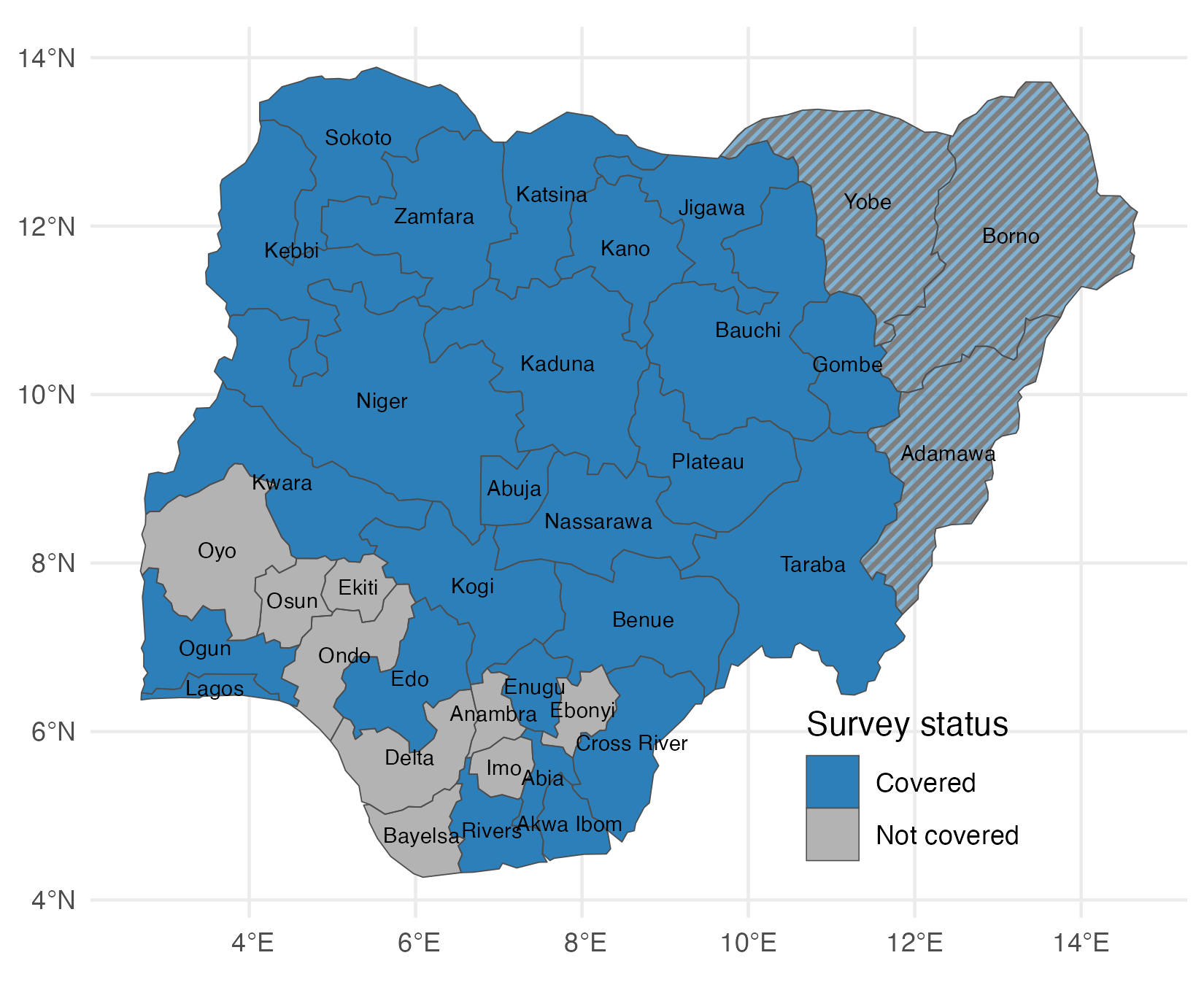}
    \caption{{Survey coverage across Nigerian states.} States shown in blue correspond to regions with standard survey coverage based on a 90-day rolling window. States in grey indicate regions without survey coverage during the study period. States with diagonal shading (Borno, Yobe, and Adamawa) are covered by surveys but with a higher-frequency 30-day rolling window, reflecting a different data collection protocol.}
    \label{fig: Nigeria map}
\end{figure}

Figure~\ref{fig: Nigeria map} illustrates the spatial distribution of survey coverage across Nigerian states. 
The majority of states are covered by surveys using a 90-day rolling window, including Sokoto, Kebbi, Zamfara, Katsina, Kano, Jigawa, Bauchi, Gombe, Kaduna, Niger, Plateau, Taraba, Kogi, Benue, Enugu, Cross River, Akwa Ibom, Rivers, Edo, Ogun, Lagos, and the Federal Capital Territory (Abuja).  A subset of states is not covered by surveys during the study period, namely Oyo, Osun, Ekiti, Ondo, Delta, Bayelsa, Anambra, Imo, Abia, and Ebonyi. In addition, three north-eastern states, Borno, Yobe, and Adamawa, are covered under a different data collection protocol using a 30-day rolling window instead of the 90-day window. Due to this difference in temporal aggregation and the associated inconsistency with the rest of the dataset, we do not include these three states in the main analysis.

\section{Spatial cross validation scheme}

To evaluate model performance under spatially heterogeneous data coverage, we designed a spatially balanced cross-validation strategy. Cross-validation folds are constructed by holding out entire regions rather than individual observations, which preserves the complete grid structure required for Kronecker-based inference on the training set.

\begin{figure}[!htbp]
    \centering
    \includegraphics[width=0.8\linewidth]{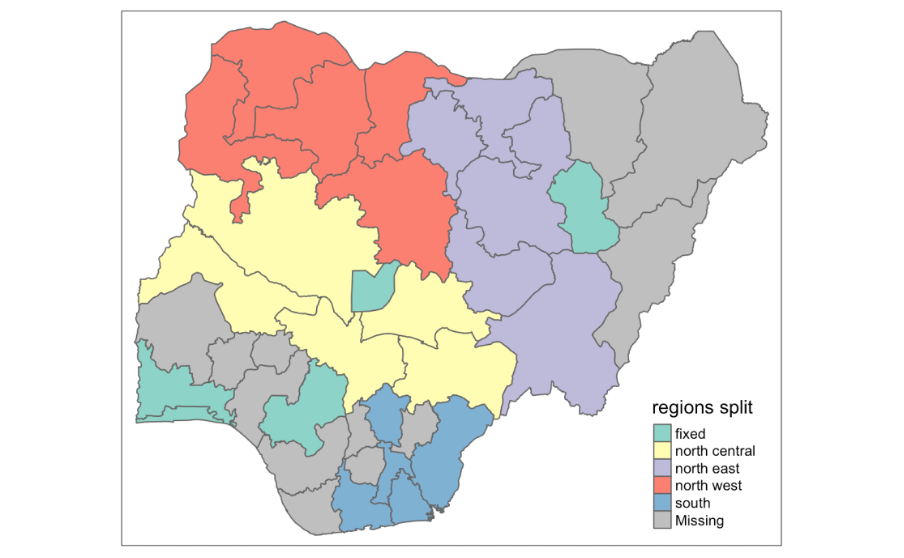}
    \caption{{Spatial grouping used for cross-validation in Nigeria.} States are partitioned into geographically coherent groups (colours). Cross-validation folds are constructed by selecting one state from each group per fold, ensuring that each test set is spatially balanced across the country. Green regions denote states that were always retained in the training set.}
    \label{fig: Nigeria CV map}
\end{figure}

For Nigeria, states were first partitioned into geographically coherent groups (Figure~\ref{fig: Nigeria CV map}). Cross-validation folds were then constructed by selecting one state from each group per fold, resulting in five folds with geographically distributed test sets. This ensures that each fold contains a balanced representation of different parts of the country, allowing for a more reliable assessment of model generalisation across space.
In addition, a subset of states (shown in green in Figure~\ref{fig: Nigeria CV map}) was always retained in the training set and never included in either the test or validation splits. These states correspond to regions with particular characteristics, such as the capital or geographically isolated areas, for which holding them out would lead to unrepresentative evaluation settings.

Within each outer fold, the selected states form the test set. Hyper-parameter tuning was performed within the training data using an inner validation split. For the XGBoost model, this consisted of splitting the training data into training and validation subsets to select optimal hyper-parameters. A similar nested procedure was used for the Gaussian process model to select the spatial length-scale parameter ($\rho_s$), with candidate values evaluated based on validation performance under root mean squared error. 

For Chad, where data coverage is more complete, regions were randomly partitioned into six folds. The same nested cross-validation procedure was applied for hyper-parameter tuning.

\section{Modelling details}

\subsection*{Kronecker-based computation with an additional non-structured regional random effect}

While direct inference for Gaussian process models scales cubically in the number of observations, requiring ${O}(n^3)$ operations and ${O}(n^2)$ storage to evaluate the likelihood and posterior quantities, Kronecker-based methods reduce the computational cost substantially by operating on lower-dimensional spatial and temporal covariance matrices, exploiting the separable spatio-temporal structure of the covariance. We follow\cite{ishida2025hierarchical} to extend their method to our setting and achive the same computational gains.\\

Given any kernel $k$ on an input domain and observed points $z_1,\ldots,z_n$, its centred version is
\begin{align}
k^{(c)}(z,z') \;=\;& k(z,z')
-\frac{1}{n}\sum_{i=1}^n k(z,z_i)
-\frac{1}{n}\sum_{i=1}^n k(z',z_i)
+\frac{1}{n^2}\sum_{i=1}^n\sum_{j=1}^n k(z_i,z_j).
\end{align}
For the spatio-temporal Gaussian process model with region-specific random effects, the marginal covariance of the vectorised response can be written as
\begin{equation}
\mathbf{K}_y
=
\alpha_0^2\left(
\mathbf{J}_{n_s} \otimes \mathbf{J}_{n_t}
+\mathbf{K}_s \otimes \mathbf{J}_{n_t}
+ \mathbf{J}_{n_s} \otimes \mathbf{K}_t
+ \mathbf{K}_s \otimes \mathbf{K}_t
\right)
+ \sigma_u^2 \left(\mathbf{I}_{n_s} \otimes \mathbf{J}_{n_t}\right)
+ \sigma_y^2 \left(\mathbf{I}_{n_s} \otimes \mathbf{I}_{n_t}\right),\label{eq: variance-covariance matrix full}
\end{equation}
where $\mathbf{K}_s$ and $\mathbf{K}_t$ are the centred spatial and temporal Gram matrices, $\mathbf{J}_m = \mathbf{1}_m \mathbf{1}_m^\top$, and $n_s,n_t$ denote the numbers of regions and time points, respectively. This extends the additive covariance used in the hierarchical ANOVA setting in \cite{ishida2025hierarchical} by adding the region-level random effect term $\sigma_u^2(\mathbf{I}_{n_s}\otimes \mathbf{J}_{n_t})$.\\

The key observation is that centred Gram matrices admit a special eigendecomposition. In particular, Proposition 2 in \cite{ishida2025hierarchical} shows that any centred Gram matrix $\mathbf{K}^{(c)}$ can be written as
\[
\mathbf{K}^{(c)} = \mathbf{Q}^{(c)} \boldsymbol{\Lambda}^{(c)} \mathbf{Q}^{(c)\top},
\]
where the first column of $\mathbf{Q}^{(c)}$ is proportional to the all-ones vector, and $\boldsymbol{\Lambda}^{(c)}$ is diagonal with first diagonal entry equal to zero. From this and Remark 1 in the same paper, it follows that
\[
\mathbf{J}_{n_s} = \mathbf{Q}_s\, \mathbf{A}_{n_s} \mathbf{Q}_s^\top,
\qquad
\mathbf{J}_{n_t} = \mathbf{Q}_t\, \mathbf{A}_{n_t} \mathbf{Q}_t^\top,
\]
where $\mathbf{Q}_s$ and $\mathbf{Q}_t$ are the orthonormal eigenvector matrices of $\mathbf{K}_s$ and $\mathbf{K}_t$, and
\[
\mathbf{A}_{n_s} = \mathrm{diag}(n_s,0,\ldots,0),
\qquad
\mathbf{A}_{n_t} = \mathrm{diag}(n_t,0,\ldots,0).
\]
This is exactly the same mechanism used in the proof of Theorem 1 for hierarchical ANOVA kernels in \cite{ishida2025hierarchical}. 
Using
\[
\mathbf{K}_s = \mathbf{Q}_s  \boldsymbol{\Lambda}_s \mathbf{Q}_s^\top,
\qquad
\mathbf{K}_t = \mathbf{Q}_t  \boldsymbol{\Lambda}_t \mathbf{Q}_t^\top,
\]
and the mixed-product property of Kronecker products, each term in $\mathbf{K}_y$ can be expressed in the common basis $\mathbf{Q}_s \otimes \mathbf{Q}_t$:

\begin{alignat*}{8}
\mathbf{J}_{n_s} &\otimes&\,\mathbf{J}_{n_t} 
& = 
(\mathbf{Q}_s \otimes \mathbf{Q}_t)
(\mathbf{A}_{n_s} &\otimes&\, \mathbf{A}_{n_t}&)&
(\mathbf{Q}_s \otimes \mathbf{Q}_t)^\top,\\
\mathbf{K}_s &\otimes&\, \mathbf{J}_{n_t}
&=
(\mathbf{Q}_s \otimes \mathbf{Q}_t)
(\boldsymbol{\Lambda}_s &\otimes&\, \mathbf{A}_{n_t}&)&
(\mathbf{Q}_s \otimes \mathbf{Q}_t)^\top,\\
\mathbf{J}_{n_s} &\otimes& \mathbf{K}_t
&=
(\mathbf{Q}_s \otimes \mathbf{Q}_t)
(\mathbf{A}_{n_s} &\otimes&\, \boldsymbol{\Lambda}_t&)&
(\mathbf{Q}_s \otimes \mathbf{Q}_t)^\top,\\
\mathbf{K}_s &\otimes& \mathbf{K}_t
&=
(\mathbf{Q}_s \otimes \mathbf{Q}_t)
(\boldsymbol{\Lambda}_s &\otimes&\,\boldsymbol{\Lambda}_t&)&
(\mathbf{Q}_s \otimes \mathbf{Q}_t)^\top,\\
\mathbf{I}_{n_s} &\otimes&\, \mathbf{J}_{n_t}
&=
(\mathbf{Q}_s \otimes \mathbf{Q}_t)
(\mathbf{I}_{n_s}&\otimes&\, \mathbf{A}_{n_t}&)&
(\mathbf{Q}_s \otimes \mathbf{Q}_t)^\top,
\end{alignat*}
and
\[
\mathbf{I}_{n_s} \otimes \mathbf{I}_{n_t}
=
(\mathbf{Q}_s \otimes \mathbf{Q}_t)
(\mathbf{I}_{n_s} \otimes \mathbf{I}_{n_t})
(\mathbf{Q}_s \otimes \mathbf{Q}_t)^\top.
\]

Therefore, we can decompose the marix $\mathbf{K}_y$ in \eqref{eq: variance-covariance matrix full} by 
\[
\mathbf{K}_y
=
(\mathbf{Q}_s \otimes \mathbf{Q}_t)\, \mathbf{D}\, (\mathbf{Q}_s \otimes \mathbf{Q}_t)^\top,
\]
with
\[
\mathbf{D}
=
\alpha_0^2
\left(
\boldsymbol{\Lambda}_s \otimes \mathbf{A}_{n_t}
+
\mathbf{A}_{n_s} \otimes \boldsymbol{\Lambda}_t
+
\boldsymbol{\Lambda}_s \otimes \boldsymbol{\Lambda}_t
\right)
+
\sigma_u^2
\left(
\mathbf{I}_{n_s} \otimes \mathbf{A}_{n_t}
\right)
+
\sigma_y^2
\left(
\mathbf{I}_{n_s} \otimes \mathbf{I}_{n_t}
\right).
\]
We note that $\boldsymbol{\Lambda}_s$, $\boldsymbol{\Lambda}_t$, $\mathbf{A}_{n_s}$, $\mathbf{A}_{n_t}$ are all diagonal with non-negative entries. As the identity matrics are diagonal with positive entry, $\mathbf{D}$ is diagonal with positive entry as well. The resulting log-determinant and linear solve are therefore obtained exactly as in the standard Kronecker approach discussed in e.g. \cite{pmlr-v37-flaxman15, wilson2014fast, saatcci2012scalable}:
\[
\log |\mathbf{K}_y|
=
\sum_{i=1}^{n_s n_t} \log \mathbf{D}_{ii},
\qquad
\mathbf{K}_y^{-1} v
=
(\mathbf{Q}_s \otimes \mathbf{Q}_t)\, \mathbf{D}^{-1}\, (\mathbf{Q}_s \otimes \mathbf{Q}_t)^\top v.
\]

Thus, the extension from the centred additive spatio-temporal GP of \cite{ishida2025hierarchical} to the present model with a region-specific random intercept is algebraically straightforward: the random effect contributes the extra diagonal Kronecker term $\sigma_u^2(\mathbf{I}_{n_s}\otimes \mathbf{A}_{n_t})$ in the rotated basis. The computational gains are the same and the estimation complexity is reduced to
${O}\left(\max\{n(n_s+n_t),\,n_s^3+n_t^3\}\right)$
and storage to ${O}(n_s^2+n_t^2)$.

\subsection*{Posterior predictive distribution} ~ Let $\mathcal{D}=\{(y_{ij},\mathbf{x}_{ij},\mathbf{s}_i,t_j): i=1,\ldots,n_s,\; j=1,\ldots,n_t\}$ denote the observed data, and let $\boldsymbol{\theta}$ collect all model parameters and hyperparameters, including regression coefficients and covariance parameters. Observations are assumed to be available on a space--time grid with $n=n_s n_t$ total observations.
Let $\mathbf{y}=(y_{11},\ldots,y_{1n_t},\ldots,y_{n_s1},\ldots,y_{n_s n_t})^\top$ denote the vectorised response, and let $\mathbf{X}$ be the corresponding design matrix without the intercept column. Conditional on the observed inputs $(\mathbf{X},\{\mathbf{s}_i\},\{t_j\})$ and parameters $\boldsymbol{\theta}$, the marginal distribution of $\mathbf{y}$ is
\begin{equation*}
\mathbf{y}\mid \mathbf{X},\{\mathbf{s}_i\},\{t_j\},\boldsymbol{\theta}
\sim 
N\!\left(
\mathbf{X}\boldsymbol{\beta},\;
\mathbf{K}
+ \sigma_u^2\,\mathbf{I}_{n_s}\otimes \mathbf{J}_{n_t}
+ \sigma_y^2\,\mathbf{I}_{n}
\right),
\end{equation*}
where $n=n_s n_t$, $\mathbf{I}_m$ is the identity matrix of dimension $m$ and $\mathbf{J}_m$ denotes an $m\times m$ matrix of ones.

The spatio-temporal covariance matrix $\mathbf{K}$ admits the Kronecker-sum representation
\begin{equation*}
\mathbf{K}
=
\left(
  \mathbf{J}_{n_s} \otimes \mathbf{J}_{n_t}
+ \mathbf{K}_s \otimes \mathbf{J}_{n_t}
+ \mathbf{J}_{n_s} \otimes \mathbf{K}_t
+ \mathbf{K}_s \otimes \mathbf{K}_t\right),
\end{equation*}
where $\mathbf{K}_s$ and $\mathbf{K}_t$ are the spatial and temporal Gram matrices with elements
$(\mathbf{K}_s)_{ii'}=k_s(\mathbf{s}_i,\mathbf{s}_{i'})$
and
$(\mathbf{K}_t)_{jj'}=k_t(t_j,t_{j'})$,
respectively. The first element $ \mathbf{J}_{n_s} \otimes \mathbf{J}_{n_t}$ takes care of the constant term $\beta_0$ and corresponds to placing prior $\beta_0\sim N(0,1)$. In practice, to avoid the variance of the normal distribution to be contrained to 1, we scale the matrix $\mathbf{K}$ by $\alpha_0^2$.

For prediction, let $\mathcal{D}^*=\{(\mathbf{x}^*_r,\mathbf{s}^*_r,t^*_r): r=1,\ldots,n^*\}$ denote a set of $n^*$ new input locations, and let $\mathbf{y}^*$ be the corresponding latent responses. Conditional on $\boldsymbol{\theta}$, the posterior predictive distribution is Gaussian,
\begin{equation*}
\mathbf{y}^* \mid \mathcal{D},\mathcal{D}^*,\boldsymbol{\theta}
\sim {N}(\mathbf{m},\mathbf{V}),
\end{equation*}
with
\begin{align*}
\mathbf{m} &= \mathbf{X}^*\boldsymbol{\beta}
+ \mathbf{K}_{* \cdot}^{\top}
\boldsymbol{\Sigma}^{-1}
(\mathbf{y}-\mathbf{X}\boldsymbol{\beta}), \\
\mathbf{V} &= \mathbf{K}_{**}
- \mathbf{K}_{* \cdot}^{\top}
\boldsymbol{\Sigma}^{-1}
\mathbf{K}_{* \cdot},
\end{align*}
where
$\boldsymbol{\Sigma}
= \mathbf{K}
+ \sigma_u^2\,\mathbf{I}_{n_s}\otimes \mathbf{J}_{n_t}
+ \sigma_y^2\,\mathbf{I}_{n}$.
Here, $\mathbf{K}_{* \cdot}$ denotes the cross-covariance matrix between prediction and observed locations with elements
$(\mathbf{K}_{* \cdot})_{r,(i,j)} = k\big((\mathbf{s}^*_r,t^*_r),(\mathbf{s}_i,t_j)\big)$,
and $\mathbf{K}_{**}$ is the covariance matrix among prediction locations with elements
$(\mathbf{K}_{**})_{rr'} = k\big((\mathbf{s}^*_r,t^*_r),(\mathbf{s}^*_{r'},t^*_{r'})\big)$.

\subsection*{Hyper parameters} \label{method:cv}~
Posterior inference for model parameters and hyperparameters is performed using Markov chain Monte Carlo with the No-U-Turn Sampler (NUTS) implemented in \texttt{Stan}. Weakly informative hyperpriors are placed on covariance parameters, with log-normal priors used for variance parameters $(\alpha_s,\alpha_{t1},\alpha_{t2})$ and an inverse-gamma prior for the temporal length-scale parameter $\rho_{t1}$. 

The spatial length-scale parameter $\rho_s$ is known to be weakly identifiable jointly with the marginal variance $\alpha_s$, as similar covariance structures can be obtained by increasing one while decreasing the other\cite{zhang2004inconsistent}. To avoid poor MCMC mixing arising from this confounding, we fix $\rho_s$ using spatial cross-validation over a predefined grid of plausible values. Candidate values for $\rho_s$ were defined over a grid spanning the range of observed inter-region distances. For Nigeria, we considered $\rho_s \in \{300, 600, 1200\}$~km, and for Chad, $\rho_s \in \{325, 650, 1300, 2600\}$~km. For each outer cross-validation fold, we evaluated these candidate values using only the training data, selecting the value that provided the best predictive performance while ensuring satisfactory MCMC convergence diagnostics.
This procedure consistently selected $\rho_s = 300$~km for Nigeria and $\rho_s = 325$~km for Chad across all folds. Given this stability, we fixed $\rho_s$ to these values in the final models.

We consider several priors for the regression coefficients $\boldsymbol{\beta}$ to assess robustness with respect to covariate selection. Specifically, we evaluate normal, ridge, and horseshoe priors, which differ in their degree of shrinkage toward zero. While the horseshoe prior offers strong variable selection properties through heavy tails, it led to slow mixing and convergence issues in our setting. We therefore adopt simpler shrinkage priors: a normal prior for the Nigerian model and a ridge prior for the Chad model, both of which achieved stable inference and comparable predictive performance with substantially improved computational efficiency. 

\section{Implementation of model comparison}

Bayesian Ridge regression is implemented using the Python library \verb|scikit-learn| with its default choices for priors and parameters.

The MLP is a fully-connected feed-forward network with three hidden layers of 128, 64 and 32 units, each followed by a ReLU activation. Dropout with rates of 0.3 and 0.5 is applied after the first and second hidden layers respectively. A final linear layer maps the 32-dimensional representation to a single scalar output. The network is trained for 200 epochs using the Adam optimizer with a learning rate of 0.001, root Mean Squared Error (rMSE) as the loss function, and a batch size of 32. We retain the checkpoint that achieves the best validation performance, which does not necessarily correspond to the final training epoch. To quantify prediction uncertainty, we employ Monte Carlo Dropout\cite{gal2016dropout}, a method that introduces stochasticity by keeping dropout activated during inference. For each prediction, we generate 100 independent forward passes with randomly deactivated neurons and construct prediction intervals from the resulting distribution. 

XGBoost is a method based on gradient boosting that sequentially combines decision trees to improve predictive accuracy. Compared to the WFP implementations\cite{martini2022machine, foini2023forecastability}, we repurpose it as a baseline in a slightly different setting, where the objective is to estimate weekly prevalence trajectories for regions that are intermittently surveyed or entirely unsurveyed during parts of the study period. Uncertainty is provided by training 100 bootstrapped models, as in the authors' implementation. The implementation follows the aforementioned works.


\section{Additional results}

\subsection*{Regression coefficient} 
\begin{figure}
    \centering

    \includegraphics[width=0.4\textheight]{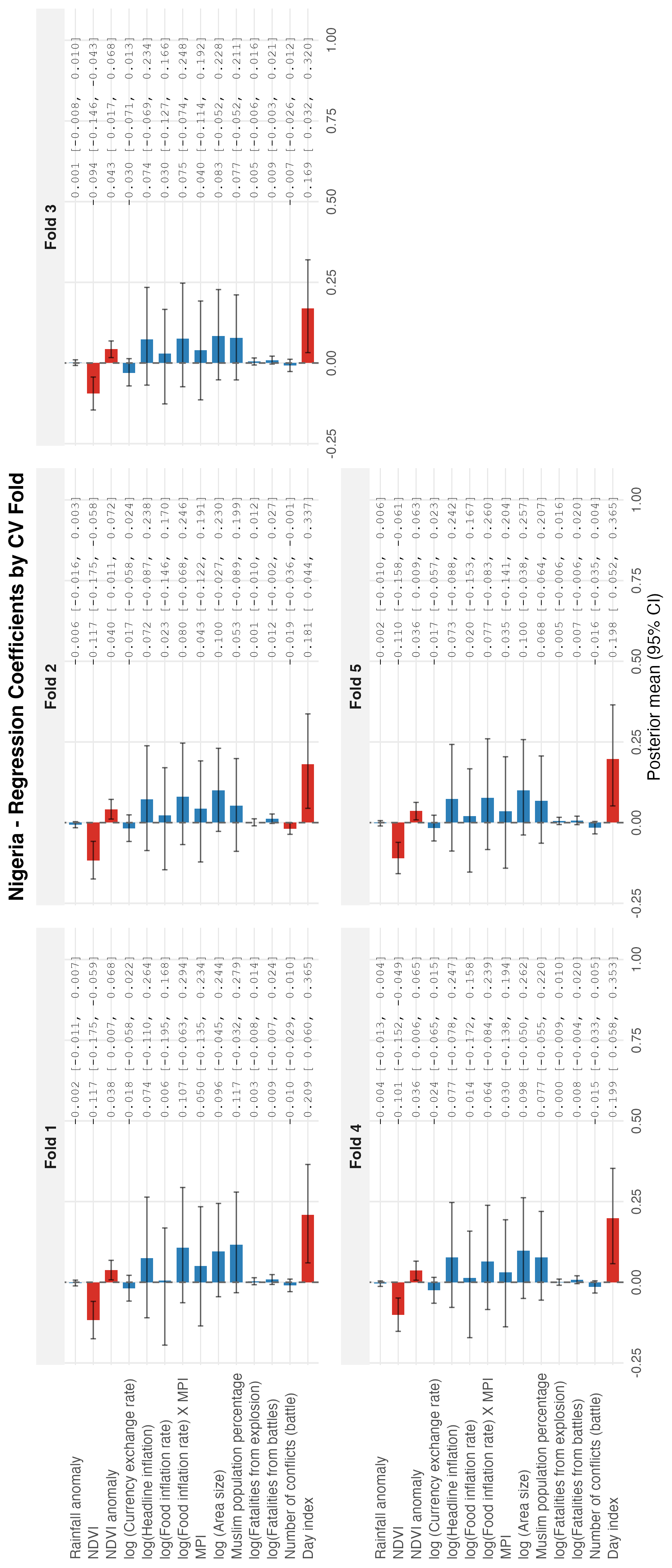}
    \vspace{10pt}
    \includegraphics[width=0.2\textheight]{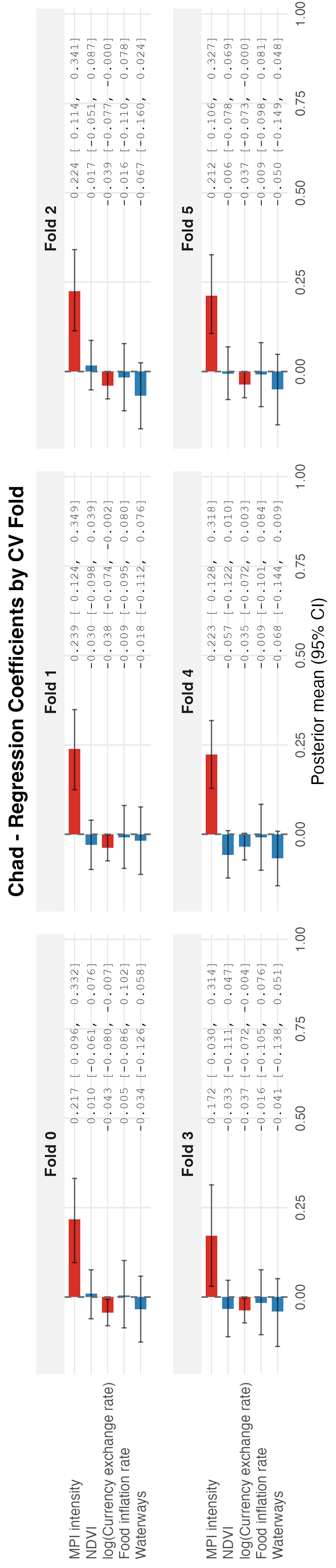}

    \caption{Estimated coefficients across cross-validation folds for Nigeria (top) and Chad (bottom). Values displayed on each bar represent the posterior mean estimates and it upper and lower 95\% credible interval.}
    \label{fig:coef_all}
\end{figure}

The top panel of Figure \ref{fig:coef_all} presents the posterior means and 95\% credible intervals of the regression coefficients for Nigeria, estimated separately for each cross-validation fold. Variables shown in red are those whose credible interval excludes zero, indicating a robust association with the outcome. Three variables are consistently identified across all five folds: NDVI value, NDVI anomaly, and the day index. The stability of these estimates across folds suggests that their effects are consistent across regions and time periods rather than driven by any particular subset of the data. The bottom panel of Figure \ref{fig:coef_all} presents the equivalent results for Chad. Two variables are consistently identified across all six folds: MPI intensity, with a positive posterior mean and the log currency exchange rate, with a negative posterior mean. This suggests that structural poverty and macroeconomic conditions are the most robust predictors of food insecurity in Chad. 

These observations deserves some nuance. The absence of a detectable linear effect does not imply that these variables are irrelevant to food security. On one hand, the spatio-temporal Gaussian process component of the model may already capture a large share of the temporal and spatial variation in the outcome, leaving little residual signal for the regression coefficients to explain. Since conflict events and climate variables exhibit strong temporal dynamics, their information may be partially absorbed by the GP prior, leading to shrinkage of the corresponding linear coefficients toward zero. On the other hand, this finding stands in interesting contrast with the broader qualitative literature on food insecurity in the Sahel, which consistently emphasises armed conflict and climate shocks as primary drivers of acute food crises \cite{crises20202020}. The discrepancy between these qualitative assessments and our quantitative results may reflect the limitations of linear parametric modelling in capturing complex, nonlinear interactions between conflict, climate, and food security outcomes.

\subsection*{Chad results}
\begin{figure}[t!]
    \centering
    \includegraphics[width=0.99\linewidth]{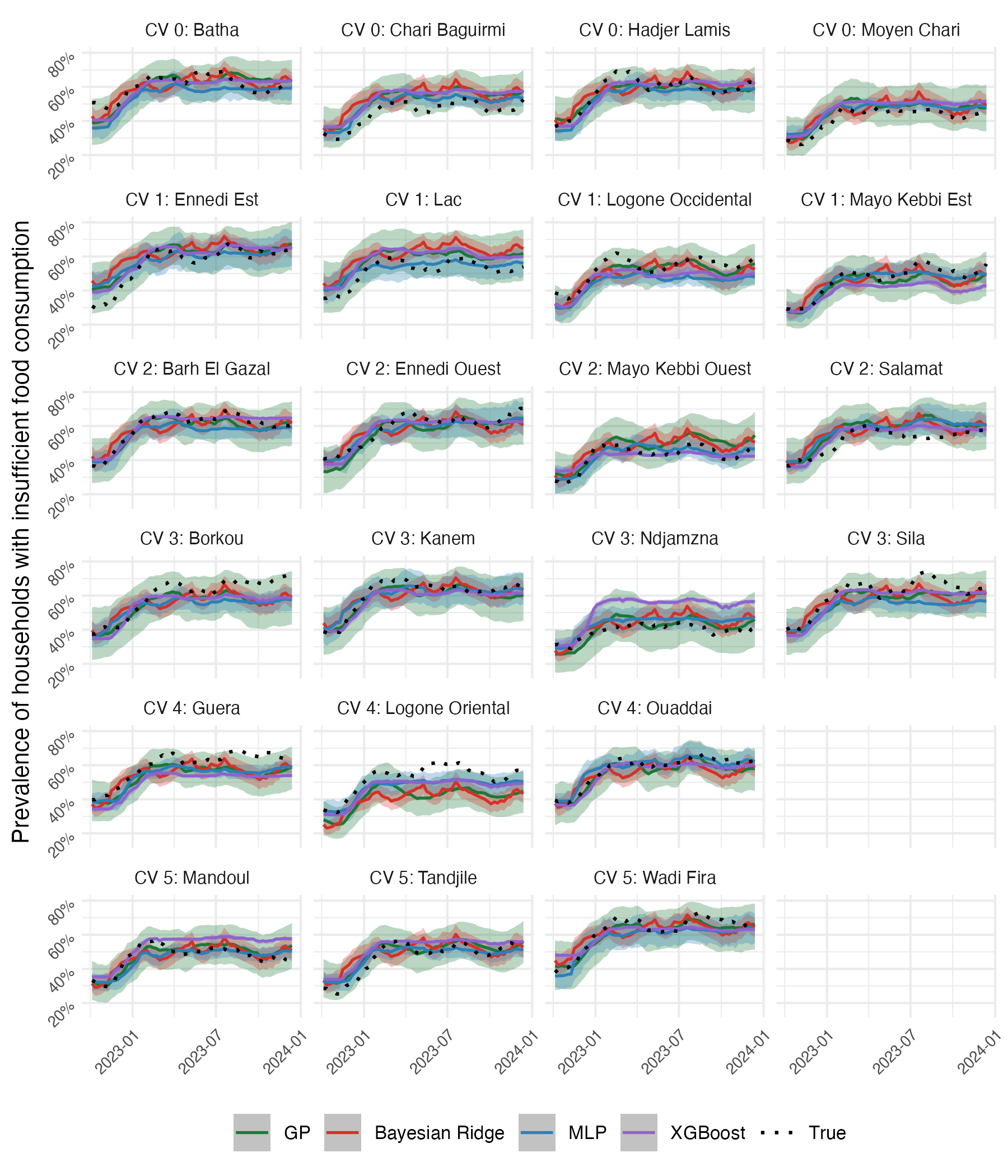}
    \caption{Prevalence of households with insufficient food consumption across states in Chad. The black dotted line denotes the observed survey-based prevalence. Coloured solid lines correspond to predictions from Gaussian Process (GP) with covariates, Bayesian Ridge, MLP, and XGBoost models. Shaded regions represent 95\% predictive intervals.}
    \label{fig: TCD prediction all regions}
\end{figure}

In \ref{fig: TCD prediction all regions} the comparative results of the four models are showed for each region in Chad. True values, point predictions and credible or prediction intervals are shown for all models.

\subsection*{Convergence diagnostics}
\begin{figure}[!htbp]
    \centering
    \includegraphics[width=\linewidth]{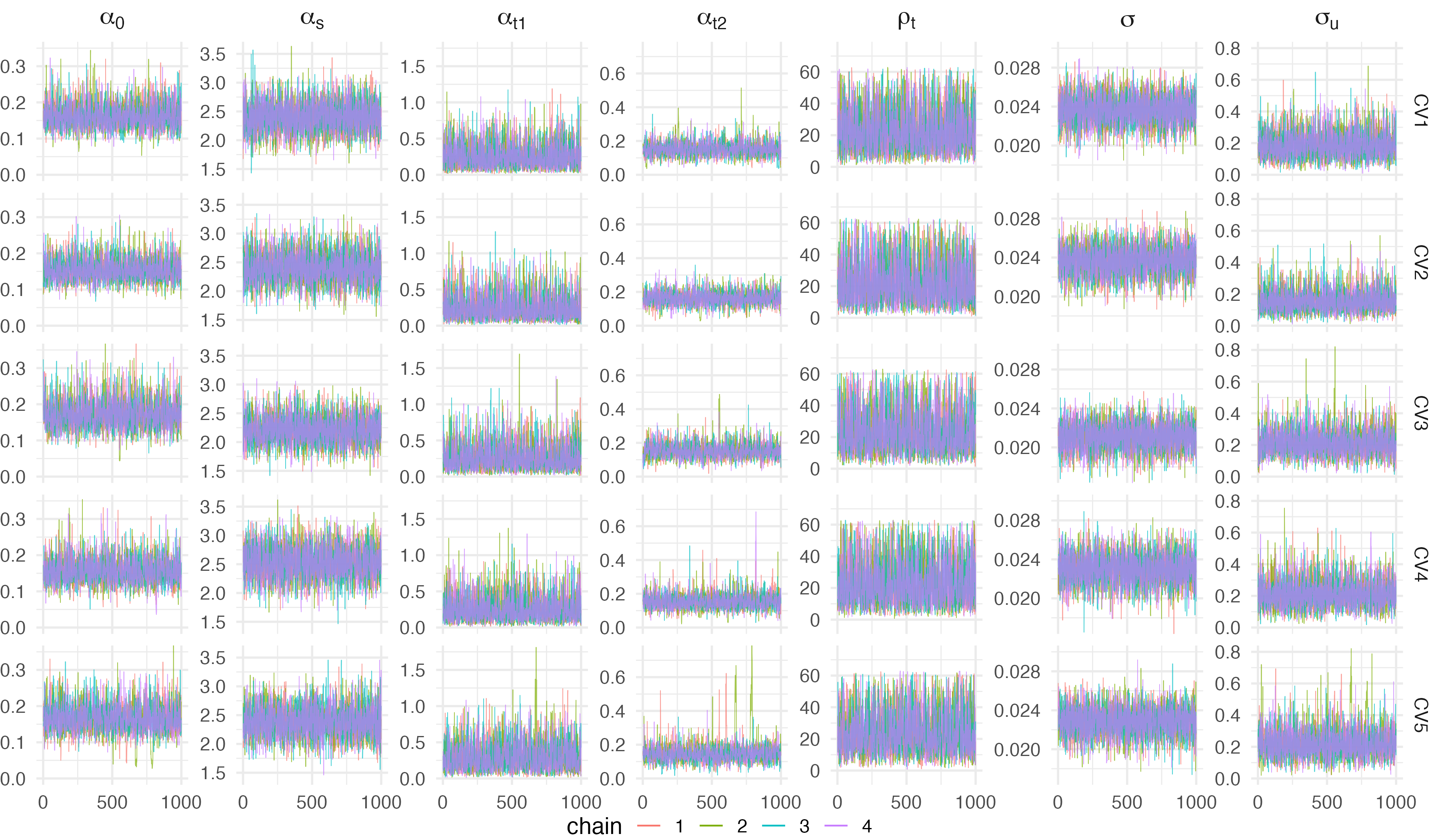}
    \caption{MCMC traceplots of the Gaussian process hyper-parameters across cross-validation folds in Nigeria.}
    \label{fig:trace_nga}
\end{figure}

\begin{figure}[!htbp]
    \centering
    \includegraphics[width=\linewidth]{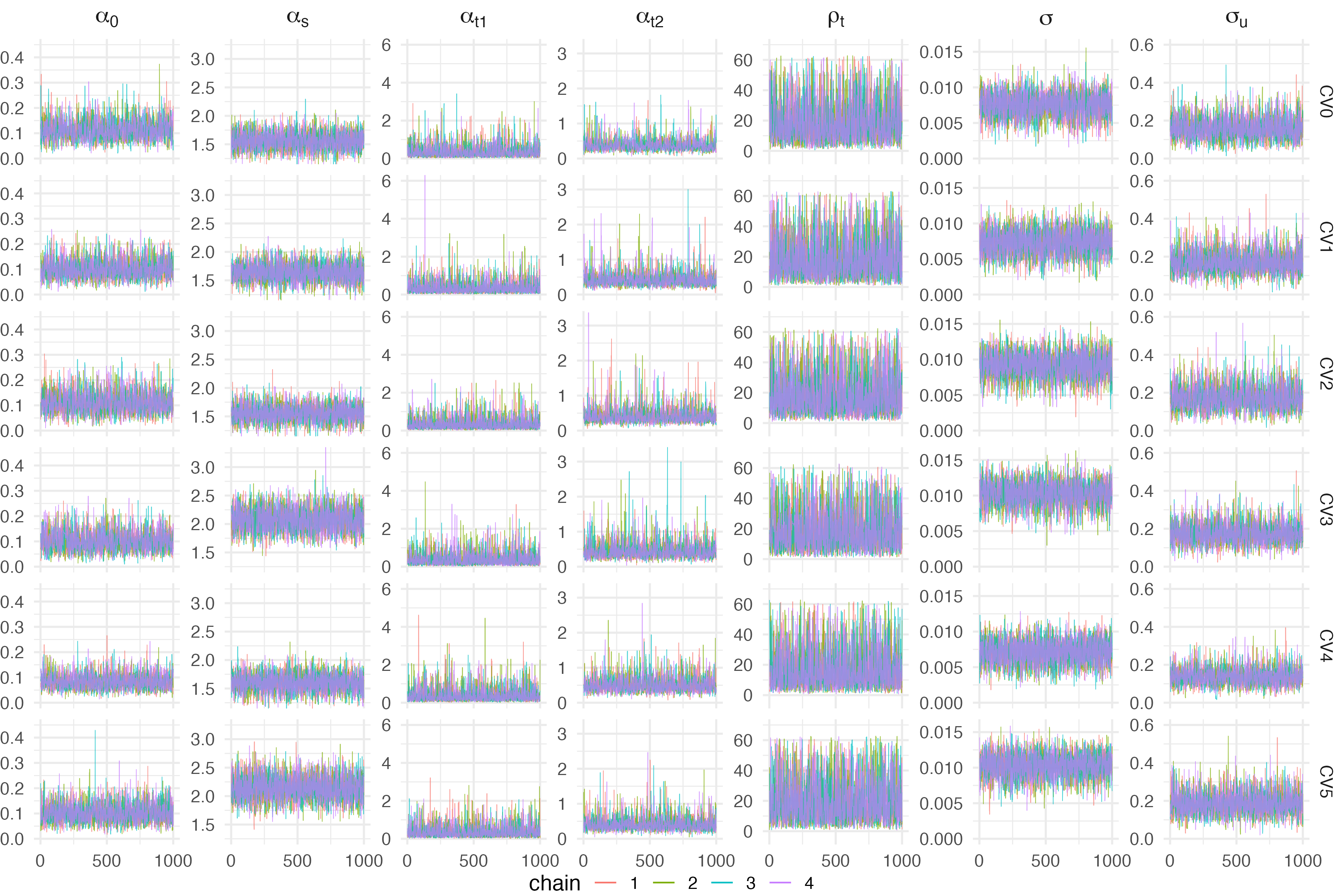}
    \caption{MCMC traceplots of the Gaussian process hyper-parameters across cross-validation folds in Chad.}
    \label{fig:trace_chad}
\end{figure}

\paragraph{Traceplots}~Figures \ref{fig:trace_nga} and \ref{fig:trace_chad} display the trace plots of the four MCMC chains across cross-validation folds for the Gaussian process hyperparameters. Each trace shows the sampled value of a given hyperparameter at every iteration of the chain. In all cases, the chains oscillate randomly around a stable value with no visible trend or drift, indicating that the sampler has mixed well and that the chains have successfully converged to the posterior distribution.

\paragraph{$\hat{R}$ statistic}~To further assess convergence, we report the maximum $\hat{R}$ statistic across all model parameters (including Gaussian process hyperparameters and regression coefficients) for each cross-validation fold and each country in Table~\ref{tab:res_cv}. All values are below $1.01$ across all folds, ranging from $1.0030$ to $1.0049$ for Nigeria and from $1.004$ to $1.0082$ for Chad. These values are well within the strict convergence threshold of $\hat{R} < 1.01$ recommended by \cite{vehtari2021rank}, confirming that the four chains have successfully converged to the posterior distribution in all settings.

\begin{table}[t!]
    \centering
\caption{Maximum $\hat{R}$ values from MCMC sampling for each fold in Nigeria and Chad, evaluated across all model parameters (Gaussian process hyperparameters and regression coefficients). Values close to 1 indicate satisfactory convergence.}
    \label{tab:res_cv}
    \begin{tabular}{lcc}
    \toprule
    Fold & Nigeria (Max $\hat{R}$) & Chad (Max $\hat{R}$) \\
    \midrule
    0 & --     & 1.0040 \\
    1 & 1.0043 & 1.0082 \\
    2 & 1.0043 & 1.0048 \\
    3 & 1.0049 & 1.0030 \\
    4 & 1.0030 & 1.0042 \\
    5 & 1.0042 & 1.0044 \\
    \bottomrule
    \end{tabular}
\end{table}


\end{document}